\documentclass[a4paper,article]{cas-sc}
\usepackage{amsmath,amsfonts}
\usepackage{array}
\usepackage{makecell}
\usepackage[caption=false,font=normalsize,labelfont=sf,textfont=sf]{subfig}
\usepackage{textcomp}
\usepackage{stfloats}
\usepackage{url}
\usepackage{verbatim}
\usepackage{graphicx}
\usepackage{nomencl}
\usepackage{etoolbox}
\usepackage{ifthen}
\usepackage{booktabs}
\usepackage[ruled]{algorithm2e}
\usepackage{appendix}
\usepackage{multirow}
\usepackage{bm}
\usepackage{makecell}
\usepackage{algpseudocode}
\allowdisplaybreaks 
\usepackage[subtle,tracking=normal]{savetrees}
\usepackage{float}

\usepackage[numbers,sort&compress]{natbib}
\usepackage{nomencl}
\makenomenclature
\renewcommand{\nomgroup}[1]{%
    \ifthenelse{\equal{#1}{A}}{\item[\textbf{Indices and Sets}]}{%
    \ifthenelse{\equal{#1}{B}}{\item[\textbf{Abbreviations}]}{%
    \ifthenelse{\equal{#1}{C}}{\item[\textbf{Parameters}]}{%
    \ifthenelse{\equal{#1}{D}}{\item[\textbf{Variables and Functions}]}{}}}}}
\usepackage{multicol}
\usepackage{framed} 
\usepackage{etoolbox} 
\usepackage{float}
\usepackage{placeins}
\usepackage{threeparttable} 

\def\tsc#1{\csdef{#1}{\textsc{\lowercase{#1}}\xspace}}
\tsc{WGM}
\tsc{QE}

\usepackage{amsthm}

\newtheorem{remark}{Remark}

\hypersetup{
    pdfauthor={Daniyaer Paizulamu, Lin Cheng, Fashun Shi, Haoyu Zheng, Pengfei He, Haiwang Zhong},
    pdftitle={Reference-Free Heterogeneous Multi-Agent Reinforcement Learning for Grid-Friendly Tie-Line Power Shaping in Industrial Microgrids},
    colorlinks=true,
    linkcolor=blue,
    citecolor=blue,
    urlcolor=blue
}

\begin{document}
\let\WriteBookmarks\relax

\renewcommand{\topfraction}{0.95}
\renewcommand{\bottomfraction}{0.90}
\renewcommand{\textfraction}{0.05}
\renewcommand{\floatpagefraction}{0.75}
\setcounter{topnumber}{3}
\setcounter{bottomnumber}{2}
\setcounter{totalnumber}{5}

\let\printorcid\relax 

\shorttitle{}
\shortauthors{}

\ExplSyntaxOn
\cs_gset:Npn \__first_footerline: { }

\cs_gset:Npn \__first_head:
{
  \parbox[t]{\textwidth}
  {
    \rule{\textwidth}{0pt}
  }
}

\cs_gset:Npn \__cas_head:
{
  \parbox[t]{\textwidth}
  {
    \rule{\textwidth}{0pt}
  }
}

\cs_gset:Npn \__first_foot:
{
  \parbox[t]{\textwidth}
  {
    \rule{\textwidth}{.2pt}\\
    \sffamily\small\makebox[\textwidth][r]{Page~\thepage{}~of~\lastpage}
  }
}

\cs_gset:Npn \__cas_foot:
{
  \parbox[t]{\textwidth}
  {
    \rule{\textwidth}{.2pt}\\
    \sffamily\small\makebox[\textwidth][r]{Page~\thepage{}~of~\lastpage}
  }
}
\ExplSyntaxOff
\pagestyle{cas}

\title[mode = title]{Reference-Free Heterogeneous Multi-Agent \\ Reinforcement Learning for Grid-Friendly Tie-Line Power \\ Shaping in Industrial Microgrids}

\author[1]{Daniyaer Paizulamu}
\credit{Writing -- original draft, Writing -- review and editing, Visualization, Validation, Software, Methodology, Formal analysis, Data curation, Conceptualization}

\author[1]{Lin Cheng}
\credit{Methodology, Supervision, Funding acquisition}

\author[1]{Fashun Shi}
\cormark[1]
\ead{shi_fashun@tsinghua.edu.cn}
\credit{Writing -- review and editing, Methodology, Supervision}

\author[1]{Haoyu Zheng}
\credit{Data curation, Visualization}

\author[1]{Pengfei He}
\credit{Data curation, Visualization}

\author[1]{Haiwang Zhong}
\credit{Writing -- review and editing, Methodology, Supervision}

\address[1]{Department of Electrical Engineering, Tsinghua University, Beijing 100084, China}

\cortext[1]{Corresponding author}

\begin{abstract}
Tie-line power (TLP) shaping is a key requirement for the grid-friendly operation of industrial microgrids (IMGs). This paper studies the coordination of multi-timescale heterogeneous adjustable resources in a steel IMG to shape a grid-friendly TLP trajectory considering multiple objectives. A sequential heterogeneous-agent coordination (SHAC) framework is proposed, where process loads, hydrogen storage, and battery storage are modeled as functionally heterogeneous agents with cross-role observations, asynchronous decision intervals, role-specific rewards and critics. This design captures the heterogeneous temporal effects of different resources on the TLP trajectory and alleviates ambiguous credit assignment and weak inter-agent coordination. To ensure feasible real-time execution, process-knowledge-based action masking and feasibility projection are embedded into policy execution, and a role-aware multi-timescale actor--critic training scheme is developed for agents with different action structures and decision intervals. Numerical studies using real renewable generation and electricity market data show that SHAC effectively eliminates the dependence on predefined reference trajectories and enables adaptive 1-min online decision-making, achieving zero production failures with an average computational time of only 0.4 ms per step. Compared with the original operation, SHAC reduces the total grid purchase cost, contract-demand exceedance time, and cumulative ramp excess by 91.27\%, 98.64\%, and 96.91\%, respectively. These results demonstrate that the proposed framework improves the economic efficiency and grid friendliness of industrial microgrid operation while satisfying strict process-safety constraints and real-time computational requirements.
\end{abstract}


\maketitle

\section{Introduction}

\subsection{Background and motivation}

An industrial microgrid (IMG) is a local energy system that coordinates industrial process loads, renewable energy sources (RES), energy storage, and the connection to the utility grid. Steel industrial microgrids have been studied to coordinate steel production, storage resources, and distributed energy resources under renewable uncertainty \cite{xi2026industrialgreenmicrogrids}. Compared with conventional microgrids, IMGs are characterized by intensive electricity consumption, strict production constraints, and highly variable power profiles. Therefore, grid-friendly IMG operation is required to coordinate production economy and grid interaction. From the active power perspective, this requirement is mainly reflected in the tie-line power (TLP), which should reduce high cost grid purchase, avoid excessive contract demand, improve local RES utilization, and suppress rapid power variations. 

Process loads provide an important source of operational flexibility for IMGs, since their production timing and operating power can be adjusted within process feasible boundaries. Existing studies have shown that flexible electric arc furnace (EAF) scheduling can reduce electricity costs, and ladle furnace (LF) refining processes can provide quantifiable load flexibility for industrial demand response \cite{zhang2025industrialflexibility}. However, short-process steelmaking loads are constrained by batch production, non-interruptible operations, and inter-stage coupling, and therefore cannot fully follow RES fluctuations. Hybrid storage can complement process load flexibility by absorbing renewable surplus and compensating production induced power variations. In particular, hydrogen storage is suitable for slow energy buffering, whereas battery storage can provide fast residual regulation \cite{su2023,giovanniello2023hybrid}. Given the distinct operational characteristics of process loads, hydrogen storage, and battery storage, designing an effective coordinated decision framework to achieve grid-friendly TLP shaping is significant but challenging.

\newpage

\nomenclature[A]{$t,\ \tau$}{Indices of operation periods and decision epochs}
\nomenclature[A]{$p,\ q,\ i$}{Indices of HS efficiency segments, storage types, and agents}
\nomenclature[A]{$\ell,\ k,\ s,\ r$}{Indices of production lines, batches, process units, and adjacent stages}
\nomenclature[A]{$\mathcal{T},\ \mathcal{T}^{\mathrm{M}},\ \mathcal{I}^{i}$}{Sets of scheduling periods, minute-level periods, and valid decision epochs}
\nomenclature[A]{$\mathcal{L},\ \mathcal{K}_{\ell},\ \mathcal{S}$}{Sets of production lines, batches, and process units}
\nomenclature[A]{$\Omega_P,\ \mathcal{A}$}{Set of HS efficiency segments and adjacent process-stage pairs}
\nomenclature[A]{$\mathcal{N},\ \mathcal{X},\ \mathcal{O}^{i},\ \mathcal{A}^{i}$}{Agent set, global state space, observation space, and action space}
\nomenclature[A]{$\mathcal{F}_{\ell,s,\tau}$}{Active-frontier set of PL scheduling candidates}

\nomenclature[B]{IMG}{Industrial microgrid}
\nomenclature[B]{TLP}{Tie-line power}
\nomenclature[B]{RES}{Renewable energy sources}
\nomenclature[B]{PL}{Process load}
\nomenclature[B]{HESS}{Hybrid energy storage system}
\nomenclature[B]{BS, HS}{Battery storage and hydrogen storage}
\nomenclature[B]{EAF, LF, CC}{Electric arc furnace, ladle furnace, and continuous caster}
\nomenclature[B]{MARL}{Multi-agent reinforcement learning}
\nomenclature[B]{SHAC}{Sequential heterogeneous-agent coordination}
\nomenclature[B]{PPO}{Proximal policy optimization}
\nomenclature[B]{GAE}{Generalized advantage estimation}
\nomenclature[B]{SOC}{State of charge}
\nomenclature[B]{SoW}{State of work}

\nomenclature[C]{$\Delta t,\ \Delta t^{\mathrm{PL}},\ \Delta t^{\mathrm{B}},\ \Delta t^{\mathrm{H}}$}{System, PL, BS, and HS time intervals}
\nomenclature[C]{$H_s,\ H_{\mathrm{H}}$}{Forecast horizon and HS residual-evaluation horizon}
\nomenclature[C]{$R^{\max}$}{Maximum acceptable minute-level TLP ramping rate}
\nomenclature[C]{$D^{\mathrm{dem}},\ \mu^{\mathrm{dem}}$}{Contract demand limit and contract demand price multiplier}
\nomenclature[C]{$\overline{P}_t^{\mathrm{RES}}$}{Available renewable power}
\nomenclature[C]{$\underline{\eta}^{\mathrm{RES}},\ \overline{\eta}^{\mathrm{ex}}$}{Minimum RES accommodation ratio and maximum grid-export ratio}
\nomenclature[C]{$c_t^{\mathrm{e}},\ \widehat c_{\tau|t}^{\mathrm{e}},\ \widehat P_{\tau|t}^{\mathrm{RES}}$}{Real-time price, forecasted price, and forecasted renewable generation}
\nomenclature[C]{$\underline{E}^{q},\ \overline{E}^{q}$}{Lower and upper energy bounds of storage \(q\)}
\nomenclature[C]{$\overline{P}^{\mathrm{B}},\ \overline{P}^{\mathrm{H,c}},\ \overline{P}^{\mathrm{H,d}}$}{BS power limit and HS charging/discharging power limits}
\nomenclature[C]{$\eta^{\mathrm{B,c}},\ \eta^{\mathrm{B,d}},\ \eta^{\mathrm{B,s}}$}{BS charging efficiency, discharging efficiency, and self-discharge rate}
\nomenclature[C]{$\alpha_p^{\mathrm{c}},\ \beta_p^{\mathrm{c}},\ \alpha_p^{\mathrm{d}},\ \beta_p^{\mathrm{d}}$}{HS piecewise affine efficiency coefficients}
\nomenclature[C]{$\underline{P}^{s},\ \overline{P}^{s},\ E_{\ell,k}^{s}$}{Power limits and energy requirement of process unit \(s\)}
\nomenclature[C]{$\underline{D}^{s},\ \overline{D}^{s}$}{Minimum and maximum processing durations}
\nomenclature[C]{$\underline{W}^{s,r},\ \overline{W}^{s,r}$}{Minimum transfer time and maximum waiting time}
\nomenclature[C]{$D^{\mathrm{EAF,gap}},\ N^{\mathrm{tar}}$}{Minimum EAF inter-batch interval and daily production target}
\nomenclature[C]{$\gamma,\ \lambda_i,\ \epsilon_{\pi}$}{Discount factor, GAE parameter, and PPO clipping parameter}
\nomenclature[C]{$c^{q},\ c^{\mathrm{ramp}},\ c^{\mathrm{H,res}}$}{Storage throughput cost, TLP ramping penalty coefficient, and HS residual penalty coefficient}
\nomenclature[C]{$\rho^{\mathrm{h}},\ c^{\mathrm{ms}},\ c^{\mathrm{soc}}$}{Revenue per heat, production shortfall penalty, and terminal SOC penalty}

\nomenclature[D]{$P_t^{\mathrm{TL}},\ P_t^{\mathrm{RES}}$}{Tie-line power and accommodated renewable power}
\nomenclature[D]{$\eta^{\mathrm{RES}},\ \eta^{\mathrm{ex}}$}{Local RES accommodation ratio and grid-export energy ratio}
\nomenclature[D]{$P_t^{\mathrm{PL}},\ P_{\ell,k,t}^{s}$}{Total PL power and processing power}
\nomenclature[D]{$x_{\ell,k,t}^{s},\ \mathrm{SoW}_{\ell,k,t}^{s}$}{Processing-status variable and completion progress}
\nomenclature[D]{$T_{\ell,k}^{s,\mathrm{st}},\ T_{\ell,k}^{s,\mathrm{ed}}$}{Start and end time slots of a batch on unit \(s\)}
\nomenclature[D]{$E_t^{\mathrm{B}},\ E_t^{\mathrm{H}}$}{Energy states of BS and HS}
\nomenclature[D]{$P_t^{q,\mathrm{c}},\ P_t^{q,\mathrm{d}},\ u_t^{q,\star}$}{Charging power, discharging power, and executable signed storage power}
\nomenclature[D]{$\eta_t^{\mathrm{H,c}},\ \eta_t^{\mathrm{H,d}}$}{Power-dependent HS charging and discharging efficiencies}
\nomenclature[D]{$x_t,\ o_t^{i},\ \boldsymbol{\chi}_t^{i}$}{Global state, local observation, and auxiliary feature vector}
\nomenclature[D]{$a_{\tau}^{\mathrm{PL}},\ a_{\tau}^{x,\mathrm{PL}},\ \boldsymbol{\rho}_{\tau}^{\mathrm{PL}}$}{PL action, discrete connection action, and normalized PL power action}
\nomenclature[D]{$\rho_t^{\mathrm{B}},\ \rho_{\tau}^{\mathrm{H}}$}{Normalized signed actions of BS and HS}
\nomenclature[D]{$\widetilde{\mathbf a}_t,\ \mathbf a_t^\star$}{Raw joint action and executable joint action}
\nomenclature[D]{$\mathcal{S}_t(\cdot)$}{Safety-aware action processor}
\nomenclature[D]{$\boldsymbol{\pi}_{\theta},\ V_{\psi^i}^{i},\ \varrho_{\tau}^{i}$}{Joint policy, role-specific critic, and PPO policy ratio}
\nomenclature[D]{$\delta_n^i,\ \widehat{A}_{\tau_n^i}^{i},\ \widehat{R}_{\tau_n^i}^{i}$}{Temporal-difference error, advantage estimate, and return target}
\nomenclature[D]{$r_t^{i},\ C_t^{\mathrm{grd}},\ C_t^{\mathrm{ramp}},\ C_t^{q,\mathrm{sto}}$}{Role-specific reward, grid cost, ramping cost, and storage throughput cost}
\nomenclature[D]{$C_t^{\mathrm{H,res}},\ C_t^{\mathrm{loss}},\ R_t^{\mathrm{fin}}$}{HS residual penalty, process loss cost, and production completion revenue}
\begin{framed}
\begin{multicols}{2}
\footnotesize
\printnomenclature
\end{multicols}
\end{framed}

\subsection{Literature review}

(1) Flexible scheduling of industrial process loads

Existing studies mainly exploit the operational flexibility of process loads from the perspectives of industrial demand response and production scheduling. Industrial facilities have been modeled as price-responsive production systems, where state-task networks, mixed-integer programming, and real-time decision models are used to coordinate production tasks with electricity price signals for reducing energy costs and enhancing demand response capability \cite{PL1,PL2,PL3}. For energy-intensive discrete industrial processes, device commitment, process sequence, batch processing, and production coupling constraints have been further incorporated, with typical applications in electric arc furnace steelmaking, refinery operations, metal casting, air separation, and cement production \cite{PL5,PL6,PL7}. For the short process steelmaking system considered in this paper, flexible EAF scheduling is investigated in \cite{PL4} to reduce electricity costs, the flexibility of the LF refining process is quantified as cuttable loads in \cite{PL8}, hydrogen-based steel production is coordinated with renewable energy supply in \cite{PL9} to improve low-carbon production economy; and risk-aware energy management of steel industrial microgrids with energy storage is studied in \cite{PL10}. In addition, continuous modeling is introduced in \cite{PL11} to improve the computational efficiency of discrete industrial process scheduling, providing a new modeling approach for fast optimization of complex process loads. However, existing studies still have two main limitations: \textit{(i)} the grid friendliness of the TLP trajectory has not been treated as a core objective. \textit{(ii)} Optimization-based methods are difficult to directly apply to minute-level online decision-making due to discrete variables, while data-driven methods have not sufficiently embedded process safety knowledge.

(2) Coordination of flexibility resources for grid-friendly TLP shaping

Existing studies have improved microgrid TLP characteristics through flexible load regulation \cite{TLP1}, storage regulation \cite{TLP2}, and coordinated operation of flexibility resources \cite{TLP3}. One class of methods constrains the power exchange between microgrids and the main grid by constructing TLP references or scheduling plans. For instance, flat TLP scheduling control is adopted in \cite{TLP4} to coordinate renewable generation and storage units, thereby enabling grid-connected microgrids to form smoother grid-interaction power profiles according to dispatch requirements. Another class of methods exploits the dynamic complementarity between flexible loads and storage for power smoothing. Demand response and battery storage are coordinated in \cite{TLP5} to share TLP fluctuations across different frequency ranges and reduce storage capacity requirements. Load shifting and storage dynamic response are combined in \cite{TLP6} to suppress microgrid TLP fluctuations, and this idea is further extended to multi-area load coordination and storage control in \cite{TLP7}. In addition, rolling optimization and multi-timescale control have been used to coordinate adjustable resources with different response speeds. Flexible loads are modeled as virtual storage and coordinated with hybrid storage for grid-connected power fluctuation smoothing in \cite{TLP8}. Hydrogen storage and battery storage are incorporated into a multi-timescale model predictive control framework in \cite{TLP9} to smooth the output power of hybrid microgrids. However, existing studies still have two main limitations: \textit{(i)} most methods mainly regulate conventional flexible loads or storage resources, without fully addressing the strong impact characteristics of process loads. \textit{(ii)} Existing methods mainly rely on predefined TLP references, filtering targets, day-ahead schedules, or rolling tracking frameworks, rather than providing a reference-free online TLP shaping method.

(3) MARL for microgrid multi-resource coordination

MARL has been increasingly applied to the coordinated control of loads, storage, and multi-resource in microgrids. For discrete manufacturing systems, production machines and processing tasks are modeled as multiple agents, and MARL is used to coordinate production scheduling and demand response for reducing electricity costs while maintaining production completion \cite{MARL1}. A similar framework is further extended to heavy industries with discrete manufacturing systems to coordinate energy demand and operating costs in energy intensive production processes \cite{MARL2}. CTDE-based MARL has been used to coordinate multiple energy resources, with soft actor--critic and attention mechanisms incorporated to handle device coupling and operational uncertainties \cite{MARL3}. The coordinated scheduling of microgrids has also been formulated as an interactive decision making process among autonomous agents to optimize local energy management and system-level operational performance \cite{MARL4}. For heat-electricity coupled microgrids, surrogate-model-enabled MARL is adopted to reduce the dependence on accurate physical models and coordinate thermal loads, electrical loads, and storage resources \cite{MARL5}. 

For storage coordination, different storage resources in hybrid storage systems are modeled as multiple agents to accommodate short, medium, and long timescale energy regulation requirements, thereby improving renewable energy utilization and microgrid operating benefits \cite{MARL6}. Shared hydrogen storage has also been incorporated into MARL-based microgrid energy management to coordinate the interactions between multiple microgrids and long-duration storage resources \cite{MARL7}. Furthermore, hierarchical or multi-timescale MARL methods have been explored for complex energy systems, where upper-level policies make slower energy allocation or scheduling decisions, while lower-level policies conduct faster power tracking or real-time correction \cite{MARL8}. Hierarchical trust-region MARL has also been used for the collaborative optimization of multi-energy multi-microgrid systems to improve training stability and coordination efficiency \cite{MARL9}. From the algorithmic perspective, HATRPO/HAPPO mitigates the training instability caused by simultaneous policy updates in cooperative MARL through sequential policy updates \cite{MARL10}. HARL/HAML further supports heterogeneous-agent cooperation with non-parameter-sharing policies, and monotonic improvement guarantees, providing a methodological basis for cooperative learning with heterogeneous action spaces and policy structures \cite{MARL12}. However, existing studies have not sufficiently characterized the differentiated responsibilities of PL, HS, and BS in TLP smoothing.

\begin{table*}[!ht] 
\footnotesize\rmfamily
  \centering
  \begin{threeparttable}
  \caption{\rmfamily Comparison of this paper with existing studies.}
  \setlength{\tabcolsep}{0.7mm}{
      \begin{tabular*}{\textwidth}{@{\extracolsep{\fill}} l c c c c c c c}
    \toprule
    Ref.
    & \makecell{Resource\\Type}
    & Algorithm
    & \makecell{Research\\Objective}
    & \makecell{Reference-\\free}
    & \makecell{Safety-\\aware}
    & \makecell{Multi-timescale\\Coordination}
    & \makecell{Time\\Resolution}\\
    \midrule

    \cite{PL5}
    & PL+BS
    & DA scheduling+MPC
    & Ancillary service
    &
    & $\checkmark$
    & $\checkmark$
    & 2-s/1-h \\

    \cite{PL4}
    & PL
    & RTN-MIP
    & Cost reduction
    &
    & $\checkmark$
    &
    & 15-min \\

    \cite{PL9}
    & PL+HS+RES
    & iRTN-MIP
    & H$_2$-steel scheduling
    &
    & $\checkmark$
    &
    & 30-min \\

    \cite{PL10}
    & PL+BS+GT
    & DRCC-MISOCP
    & Risk-aware EM
    &
    & $\checkmark$
    &
    & 1-h \\

    \cite{PL11}
    & PL
    & cRTN-MILP
    & Efficient scheduling
    &
    & $\checkmark$
    &
    & 5-min \\

    \cite{TLP1}
    & PL
    & LQ-based OR
    & TLP smoothing
    & $\checkmark$
    &
    &
    & $<5$-min \\

    \cite{TLP2}
    & BS
    & MAD3QN
    & PCC smoothing
    &
    &
    &
    & 1-h \\

    \cite{TLP3}
    & FL+BS
    & DASP+IDRC
    & Grid-friendly EM
    &
    &
    & $\checkmark$
    & 30-min \\

    \cite{TLP4}
    & RES+BS
    & Hierarchical control
    & Flat TLP scheduling
    &
    &
    &
    & 15-min \\

    \cite{TLP5}
    & FL+BS
    & SQ+KBM
    & TLP smoothing
    &
    &
    & $\checkmark$
    & 1-min \\

    \cite{TLP8}
    & FL+HESS
    & NROS
    & PFS+economy
    &
    &
    & $\checkmark$
    & 6-min/30-min \\

    \cite{TLP9}
    & HS
    & MT-MPC
    & Output smoothing
    &
    &
    & $\checkmark$
    & 10-min/1-h \\

    \cite{MARL1}
    & PL
    & MADDPG
    & Industrial DR
    &
    &
    &
    & 1-h \\

    \cite{MARL2}
    & PL+ES
    & MAPPO+AM
    & DR+production EM
    &
    &
    &
    & 1-h \\

    \cite{MARL3}
    & Load+ES+ME
    & MADRL+CB/SAC/AM
    & Cost reduction
    &
    &
    &
    & 1-h \\

    \cite{MARL8}
    & ES+DG
    & MA-OCDDPG
    & Hierarchical EM
    &
    &
    & $\checkmark$
    & 1-h \\

    \cite{MARL9}
    & MEMGs
    & CD-EHATRPO
    & Collaborative optimization
    &
    &
    & $\checkmark$
    & 15-min \\

    This paper
    & PL+HS+BS
    & SHAC
    & Grid-friendly TLP shaping
    & $\checkmark$
    & $\checkmark$
    & $\checkmark$
    & 1/5/15-min \\

    \bottomrule
    \end{tabular*}%
    }\label{literature_review}
    \begin{tablenotes}
\item $\checkmark$: The item is considered; GT: gas turbine; ME: multi-energy resources; MEMGs: multi-energy microgrids; EM: energy management; OR: output regulation; PFS: power fluctuation smoothing; KBM: kinetic battery model; MAD3QN: multi-agent dueling double deep Q-network; DASP: day-ahead scheduling plan; IDRC: intraday rolling correction; NROS: novel rolling optimization strategy; MADDPG: multi-agent deep deterministic policy gradient; AM: attention mechanism; MA-OCDDPG: multi-agent option-critic deep deterministic policy gradient; CD-EHATRPO: collaboration-driven enhanced heterogeneous-agent trust-region policy optimization.
\end{tablenotes}
\end{threeparttable}\vspace{-0.5cm}
\end{table*}%

\subsection{Research gap}
Existing literature is summarized in Table~\ref{literature_review}. Despite the extensive literature, several critical research gaps remain insufficiently addressed.

(1) Existing studies on PL scheduling mainly focus on production economy, demand response, or ancillary service provision, while the grid friendliness of the TLP trajectory has not been treated as a core optimization objective. Therefore, it is crucial to exploit PL flexibility for improving the grid-interaction power trajectory while preserving process safety constraints.

(2) Existing TLP shaping methods mostly follow a two-stage paradigm in which a predefined TLP reference trajectory is first generated and flexible resources are then coordinated to track it. However, in short-process steelmaking IMGs, process loads are characterized by discrete batch processing, non-interruptible operation, strict temporal coupling, and strong power impacts. Predefined TLP reference trajectories may not adequately accommodate real-time renewable generation, electricity price signals, and production-state variations, which may lead to degraded economic performance, excessive storage cycling, and even compromised production feasibility. Therefore, a reference-free online TLP shaping mechanism is still needed to achieve grid-friendly TLP shaping.

(3) Current MARL-based microgrid coordination studies often place flexible resources with similar functions or comparable timescales within a unified decision-making framework, which is insufficient to characterize the physical heterogeneity and hierarchical coordination of multi-type resources in industrial microgrids. In steel IMGs, PL, HS, and BS differ not only in their decision timescales but also in their functional roles in TLP shaping, where PL reshapes production-induced load profiles, HS buffers medium-timescale energy residuals, and BS compensates minute-level residual fluctuations. Under such heterogeneity, a unified reward or parallel training structure may obscure credit assignment and weaken cross-resource coordination. Therefore, a heterogeneous coordination framework oriented toward resource heterogeneity and temporal coupling is still needed to enable effective credit assignment and coordinated TLP shaping.

\subsection{Contributions}

Motivated by the above research gaps, this paper proposes a reference-free heterogeneous MARL framework for grid-friendly TLP shaping in a steel IMG. Specifically, the main contributions are threefold:

\textbf{(1) A heterogeneous Markov decision model for reference-free TLP shaping:} Considering the significant differences among process loads, hydrogen storage, and battery storage in physical characteristics, action structures, and response timescales in short-process steelmaking industrial microgrids, this paper models the grid-friendly TLP shaping problem as a heterogeneous multi-agent sequential decision-making process. This model uniformly transforms process constraints, storage dynamics, real-time operating signals, and grid-friendly objectives into a state-driven reinforcement learning decision-making problem, providing a modeling basis for reference-free online collaborative optimization of complex industrial source--load--storage systems.

\textbf{(2) A sequential heterogeneous-agent coordination framework (SHAC):} To address the ambiguous credit assignment and weakened coordination that may occur in conventional MARL for heterogeneous resource coordination, this paper proposes a PL$\rightarrow$HS$\rightarrow$BS sequential heterogeneous coordination framework. Its core is to strictly align the sequential decision-making of agents with the physical responsibility chain of resources, and to establish conditional dependencies among heterogeneous actions through cross-role observations, role-specific rewards, role-specific critics, and asynchronous update mechanism design. This design effectively alleviates the credit assignment difficulty in heterogeneous cooperation, not only enabling efficient coordinated decision-making among heterogeneous microgrid resources, but also enhancing the dynamic collaborative shaping capability of multi-timescale resources for the TLP trajectory.

\textbf{(3) A process-knowledge-embedded safety-executable policy layer:} To address the problem that purely data-driven policies are prone to generating non-executable actions in short-process steelmaking scenarios, this paper introduces a safety-aware action processing layer between heterogeneous-agent policies and the industrial microgrid environment, which uniformly embeds complex discrete process logic and continuous physical operating boundaries into the policy execution process. Through action masking and feasibility projection, this mechanism maps the heterogeneous policy outputs of PL, HS, and BS in real time into process-feasible and physically feasible online dispatch commands, thereby enhancing engineering deployability while maintaining reference-free adaptive decision-making capability.

\subsection{Paper organization}

We organize the remainder of this paper as follows. Section~\ref{characterization} describes the configuration of IMG and characterizes the requirements of grid-friendly TLP shaping. Section~\ref{modeling} presents the equivalent models of HESS, together with the operation constraints of PLs. Section~\ref{methodology} introduces the reference-free SHAC framework for grid-friendly TLP shaping. Section~\ref{Case_studies} provides numerical case studies to evaluate the effectiveness of the proposed framework. Finally, Section~\ref{conclusion} concludes this paper.

\section{Industrial microgrid description and grid-friendly TLP characterization}~\label{characterization}

\subsection{Industrial microgrid configuration}

Industrial microgrid (IMG) refers to the local energy system deployed in an industrial park, typically comprising distributed generation, PL-dominated industrial loads, energy storage devices, an internal distribution network, an energy management system, and a grid-interconnection interface. Compared with conventional microgrids, IMGs are more strongly shaped by industrial production in their resource composition, system structure, and operating patterns. Specifically, (1) the generation side often integrates large-capacity low-carbon resources, such as wind power, photovoltaic generation, and waste-heat or waste-pressure recovery generation, rather than only small-scale distributed units. (2) The load side is dominated by high-power industrial equipment and automated production lines, exhibiting nonlinear, impulsive, and strongly process-coupled characteristics. (3) The operating scale typically ranges from tens to hundreds of MW, with possible interconnection at higher voltage levels and coverage of an entire industrial plant or park. (4) The operation of an IMG aims not only to improve local economy and supply reliability, but also to enhance local renewable energy accommodation and improve TLP interactions with the upstream distribution grid through coordinated source-load-storage scheduling.

\begin{figure}[pos=!t]
  \footnotesize\rmfamily   \setlength{\abovecaptionskip}{-0.1cm}  
    \setlength{\belowcaptionskip}{-0.1cm} 
  \begin{center}  \includegraphics[width=0.5\columnwidth]{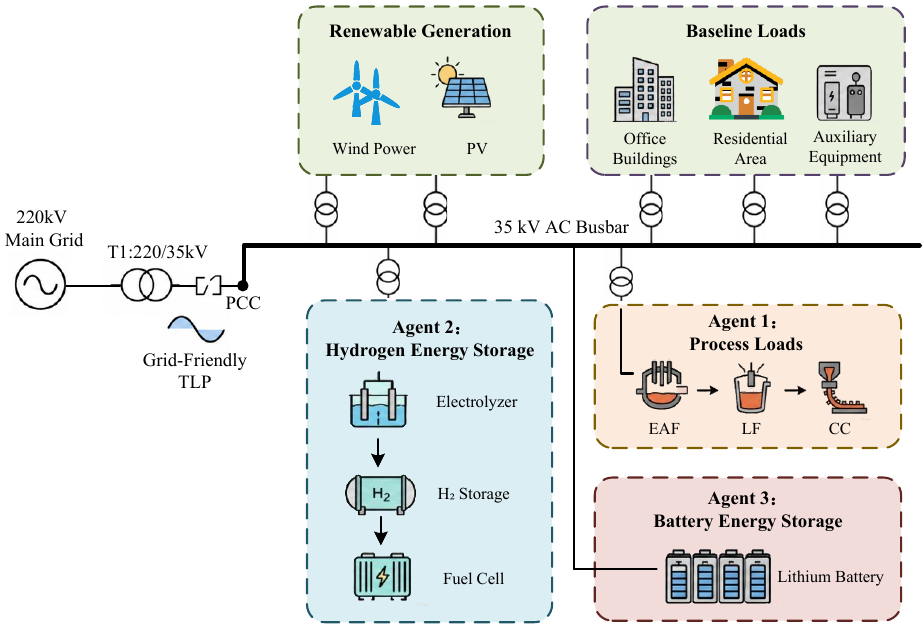}
     \caption{\rmfamily Architecture of the studied steel industrial microgrid with process loads and hybrid energy storage.}\label{IMG}
  \end{center}
  \vspace{-2.5em}
\end{figure}

As shown in Fig.~\ref{IMG}, the steel IMG studied in this paper consists of wind and photovoltaic renewable energy sources (RESs), short-process steelmaking process loads (PLs), a hybrid energy storage system (HESS), and an external grid interface. The steelmaking PLs mainly include EAFs, LFs, and continuous casters (CCs), and are constrained by batch production, inter-process coordination, equipment occupancy, and intraday production requirements. The HESS integrates hydrogen storage and battery energy storage systems, providing long- and short-duration energy shifting capabilities, respectively. This paper focuses on the coordinated dispatch of PLs and the HESS to shape a more grid-friendly tie-line power \(P_t^{\mathrm{TL}}\) in real time while satisfying production requirements and equipment operating constraints.

\subsection{Grid-friendly TLP characterization}\label{Grid-friendly_TLP_characterization}

\begin{figure}[pos=!t]
  \footnotesize\rmfamily   \setlength{\abovecaptionskip}{-0.1cm}  
    \setlength{\belowcaptionskip}{-0.1cm} 
  \begin{center}  \includegraphics[width=0.7\columnwidth]{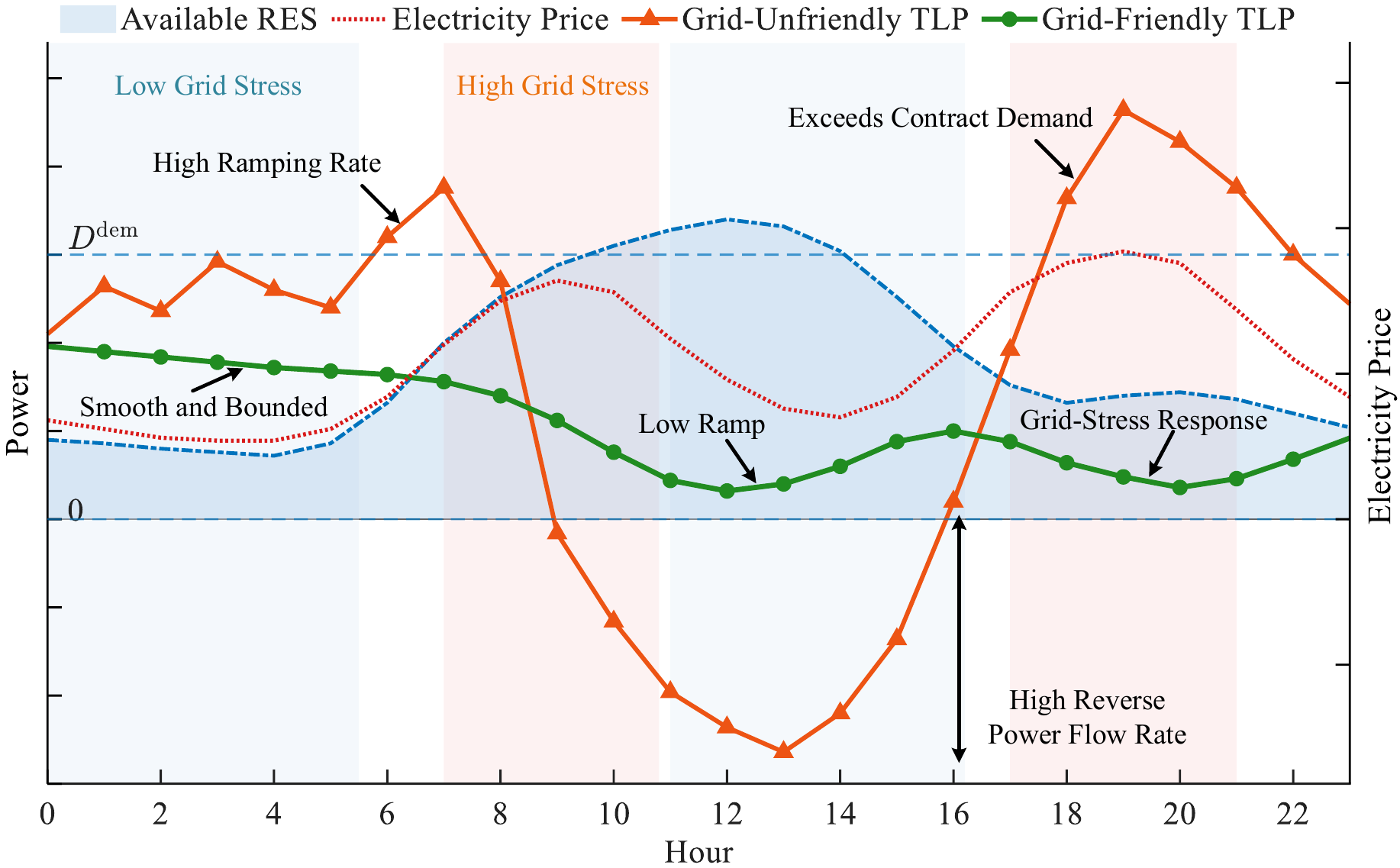}
     \caption{\rmfamily Illustrative comparison of grid-unfriendly and grid-friendly IMG tie-line power trajectories under renewable generation and electricity price signals.}\label{TLP}
  \end{center}
  \vspace{-2.5em}
\end{figure}

In this paper, a grid-friendly TLP refers to a power trajectory that attenuates the direct transmission of internal PL shocks and renewable energy fluctuations from the industrial microgrid to the upstream grid, coordinates electricity procurement with grid operational needs, and presents an observable, controllable, bounded, and coordinated interaction pattern at the grid interface. Specifically, Fig.~\ref{TLP} illustrates the contrast between grid-unfriendly and grid-friendly TLP trajectories, and a grid-friendly TLP should exhibit the following characteristics~\cite{zhinan}.

(1) \textit{Smoothness:} Smoothness requires the TLP to change gradually rather than directly transferring abrupt PL variations or renewable fluctuations to the upstream grid. This property can be characterized by the admissible ramping-rate limit:
\begin{equation}
\left|\frac{P_t^{\mathrm{TL}}-P_{t-1}^{\mathrm{TL}}}{\Delta t}\right|
\leq R^{\max}, \quad \forall t\in\mathcal{T},
\end{equation}

\noindent where \(R^{\max}\) denotes the maximum acceptable minute-level TLP ramping rate. A lower ramping rate implies a smoother grid-interaction trajectory.

(2) \textit{Local renewable accommodation and limited grid export:}
A grid-friendly IMG should prioritize the local accommodation of wind and photovoltaic generation by industrial loads and storage devices. Surplus renewable generation may be exported to the upstream grid for spot market participation, whereas the exported energy should remain limited relative to the total available renewable generation. This property can be characterized by:
\begin{subequations}
\label{eq:res_accommodation_export}
\begin{align}
&\eta^{\mathrm{RES}}=\frac{
\sum_{t\in\mathcal{T}} P_t^{\mathrm{RES}}\Delta t
}{\sum_{t\in\mathcal{T}} \overline{P}_t^{\mathrm{RES}}\Delta t}\geq
\underline{\eta}^{\mathrm{RES}},
\label{eq:res_accommodation_ratio}\\
&\eta^{\mathrm{ex}}=\frac{\sum_{t\in\mathcal{T}} [-P_t^{\mathrm{TL}}]^+\Delta t
}{\sum_{t\in\mathcal{T}} \overline{P}_t^{\mathrm{RES}}\Delta t}\leq
\overline{\eta}^{\mathrm{ex}},
\label{eq:grid_export_ratio}
\end{align}
\end{subequations}

\noindent where $\overline{P}_t^{\mathrm{RES}}$ denotes the available renewable power at time $t$, $P_t^{\mathrm{RES}}$ denotes the locally accommodated renewable power, and $[-P_t^{\mathrm{TL}}]^+$ denotes the exported power. $\underline{\eta}^{\mathrm{RES}}$ imposes the lower limit on the local renewable accommodation ratio, while $\overline{\eta}^{\mathrm{ex}}$ imposes the upper limit on the grid-export energy ratio.

(3) \textit{Bounded grid import:}
Grid-friendly TLP should remain within the admissible import capacity accepted by the upstream grid. For an IMG, this capacity can be represented by its contracted demand $D^{\mathrm{dem}}$. Accordingly, the TLP is required to satisfy:
\begin{equation}
P_t^{\mathrm{TL}}\leq D^{\mathrm{dem}},
\quad \forall t\in\mathcal{T}.
\end{equation}

(4) \textit{Grid-stress responsiveness:}
A grid-friendly TLP should be coordinated with the operating stress of the upstream grid. When the upstream grid is under high-load, high-price, or high-net-load conditions, the IMG is expected to reduce its grid import. Conversely, when grid stress is low or surplus renewable generation is available, the IMG may increase local consumption or charge storage.

\begin{remark}[Grid Export and Renewable Curtailment]
Renewable curtailment is not adopted as a direct control lever for reducing exported power in this paper. Unrestricted curtailment would reduce the grid-export ratio through the disposal of available renewable generation, rather than through the effective coordination of industrial process loads and hybrid energy storage. This treatment would obscure the attribution of grid-export reduction to the cooperative actions of heterogeneous agents. Therefore, the limited grid-export requirement is mainly achieved through coordinated production scheduling and storage charging/discharging, rather than through renewable energy disposal.
\end{remark}

\begin{remark}[Grid-Stress Responsiveness]
The real-time electricity price is adopted as a measurable proxy for the operating stress of the upstream grid. Elevated prices generally reflect stressed operating conditions associated with high demand, limited supply, or network congestion, whereas low prices indicate relatively relaxed operating conditions or abundant low-cost generation. When local renewable generation is insufficient, shifting production toward low-price periods and avoiding high-price periods reduces grid import under stressed grid conditions, thereby enabling grid-stress-responsive TLP shaping.
\end{remark}

\section{Hybrid energy storage and process loads modeling}~\label{modeling}

Considering that the net load of an industrial microgrid exhibits both medium-to-long-duration energy imbalance and short-term power fluctuations within the intraday horizon, a hybrid energy storage system composed of hydrogen storage (HS) and battery energy storage (BS) is adopted to facilitate the shaping of a smoother and more grid-friendly $P^{\mathrm{TL}}$ trajectory. The battery energy storage system mitigates minute-level power fluctuations through its fast response capability, while the hydrogen storage system regulates net-load deviations at approximately the $15$-min timescale by leveraging its longer-duration energy shifting capability.

\subsection{Equivalent hydrogen and battery storage model}~\label{HSmodeling}

The HS consists of an electrolyzer, a hydrogen tank, and a fuel cell, where electrical energy is stored through electrolytic hydrogen production during charging and released through fuel-cell generation during discharging. The conversion efficiencies of both processes are not constant but depend on the operating power. Following the power-dependent efficiency characterization in \cite{qi2025}, the equivalent HS model is formulated as follows:

\textbf{Constraints}: $\forall t\in\mathcal{T}, \forall p\in\Omega_P$
\begin{subequations}
\label{eq:HS_model}
\begin{align}
&E_{t+1}^{\mathrm{H}}
=
E_t^{\mathrm{H}}
+
\Delta t^{\mathrm{H}}
\left(
\eta_t^{\mathrm{H,c}}P_t^{\mathrm{H,c}}
-
\frac{P_t^{\mathrm{H,d}}}{\eta_t^{\mathrm{H,d}}}
\right),
\label{eq:HS_state}\\
&\eta_t^{\mathrm{H,c}}
=
\eta^{\mathrm{H,c}}\left(P_t^{\mathrm{H,c}}\right),
\quad
\eta_t^{\mathrm{H,d}}
=
\eta^{\mathrm{H,d}}\left(P_t^{\mathrm{H,d}}\right),
\label{eq:HS_efficiency_mapping}\\
&\eta^{\mathrm{H,c}}(P)
=
\alpha_p^{\mathrm{c}}P+\beta_p^{\mathrm{c}},
\quad
P\in
\left[
\underline{P}_{p}^{\mathrm{H,c}},
\overline{P}_{p}^{\mathrm{H,c}}
\right],
\label{eq:HS_charging_efficiency}\\
&\eta^{\mathrm{H,d}}(P)
=
\alpha_p^{\mathrm{d}}P+\beta_p^{\mathrm{d}},
\quad
P\in
\left[
\underline{P}_{p}^{\mathrm{H,d}},
\overline{P}_{p}^{\mathrm{H,d}}
\right],
\label{eq:HS_discharging_efficiency}\\
&\underline{E}^{\mathrm{H}}
\leq
E_t^{\mathrm{H}}
\leq
\overline{E}^{\mathrm{H}},
\label{eq:HS_energy_bound}\\
&0
\leq
P_t^{\mathrm{H,c}}
\leq
\overline{P}^{\mathrm{H,c}},
\quad
0
\leq
P_t^{\mathrm{H,d}}
\leq
\overline{P}^{\mathrm{H,d}}.
\label{eq:HS_power_bound}
\end{align}
\end{subequations}

\noindent where $\mathcal{T}$ and $\Omega_P$ denote the sets of time periods and efficiency segments, respectively. $E_t^{\mathrm{H}}$ is the HS energy state, and $P_t^{\mathrm{H,c}}$ and $P_t^{\mathrm{H,d}}$ are the charging and discharging powers of HS. $\eta_t^{\mathrm{H,c}}$ and $\eta_t^{\mathrm{H,d}}$ are the corresponding power-dependent efficiencies. Parameters $\alpha_p^{\mathrm{c}}$, $\beta_p^{\mathrm{c}}$, $\alpha_p^{\mathrm{d}}$, and $\beta_p^{\mathrm{d}}$ define the piecewise affine efficiency curves, and the active segment is determined by the realized operating power. Constraints \eqref{eq:HS_energy_bound} and \eqref{eq:HS_power_bound} impose the energy and power limits of HS.

The charging and discharging efficiencies of BS are assumed to be constant. The BS model is then formulated as follows:

\textbf{Constraints}: $\forall t\in\mathcal{T}$
\begin{subequations}
\label{eq:BS_model}
\begin{align}
&E_{t+1}^{\mathrm{B}}
=
\left(1-\eta^{\mathrm{B,s}}\right)E_{t}^{\mathrm{B}}
+
\Delta t^{\mathrm{B}}
\left(
\eta^{\mathrm{B,c}}P_{t}^{\mathrm{B,c}}
-
\frac{P_{t}^{\mathrm{B,d}}}{\eta^{\mathrm{B,d}}}
\right),
\label{eq:BS_state} \\
&0
\leq
P_{t}^{\mathrm{B,c}}
\leq
\overline{P}^{\mathrm{B}}
\left(1-b_{t}^{\mathrm{B}}\right),
\label{eq:BS_charge_power} \\
&0
\leq
P_{t}^{\mathrm{B,d}}
\leq
\overline{P}^{\mathrm{B}}b_{t}^{\mathrm{B}},
\label{eq:BS_discharge_power} \\
&\underline{E}^{\mathrm{B}}
\leq
E_{t}^{\mathrm{B}}
\leq
\overline{E}^{\mathrm{B}},
\label{eq:BS_energy_bound}
\end{align}
\end{subequations}

\noindent where $E_{t}^{\mathrm{B}}$ denotes the BS energy state. $P_{t}^{\mathrm{B,c}}$ and $P_{t}^{\mathrm{B,d}}$ are the charging and discharging power of BS, respectively. $\eta^{\mathrm{B,c}}$ and $\eta^{\mathrm{B,d}}$ are the charging and discharging efficiencies, and $\eta^{\mathrm{B,s}}$ denotes the self-discharge rate. $\overline{P}^{\mathrm{B}}$ is the maximum charging/discharging power. $\underline{E}^{\mathrm{B}}$ and $\overline{E}^{\mathrm{B}}$ are the lower and upper bounds of the BS energy state. $b_{t}^{\mathrm{B}}$ is a binary variable.

\subsection{Process loads modeling}~\label{PLmodeling}

This paper focuses on the PLs in a short-process steelmaking industrial microgrid. A typical short-process steelmaking system consists of multiple parallel production lines, each of which follows a fixed technological sequence for steelmaking, refining, and continuous casting. The main process units include the EAF, LF, and CC. Specifically, the EAF melts scrap steel and performs primary steelmaking, representing the PL with the highest power level and the most pronounced power fluctuations. The LF adjusts the chemical composition and temperature of molten steel, and its operation is subject to both processing-duration requirements and inter-process coordination constraints. The CC casts the refined molten steel into billets, imposing strict requirements on upstream batch arrival times and production continuity. Since steelmaking is characterized by batch processing, the scheduling flexibility of PLs mainly arises from selecting the start times of different batches across production lines and time periods. Therefore, PLs are modeled as production-coupled loads with discrete start-up decisions, non-interruptible processing, and multi-stage temporal coupling.

\subsubsection{Process loads state equation}~\label{PLstate}

We introduce the state of work (SoW) equation to describe the processing progress of each batch at different stages. Let $\mathcal{L}$ denote the set of production lines, $\mathcal{S}=\{\mathrm{EAF},\mathrm{LF},\mathrm{CC}\}$ denote the set of process units, $\mathcal{K}_{\ell}$ denote the set of batches on production line $\ell$, and $\mathcal{T}$ denote the set of scheduling intervals. For batch $k$ on production line $\ell$, $\mathrm{SoW}_{\ell,k,t}^{s}\in[0,1]$ denotes its completion progress on unit $s$ at time $t$. The state update is given by
\begin{equation}
\mathrm{SoW}_{\ell,k,t}^{s}=\mathrm{SoW}_{\ell,k,t-1}^{s}+x_{\ell,k,t}^{s}
\frac{P_{\ell,k,t}^{s}\Delta t^{\mathrm{PL}}}{
E_{\ell,k}^{s}},
\quad
\forall \ell\in\mathcal{L},\;
k\in\mathcal{K}_{\ell},\;
s\in\mathcal{S},\;
t\in\mathcal{T},
\end{equation}

\noindent where $x_{\ell,k,t}^{s}\in\{0,1\}$ is the processing-status variable. Specifically, $x_{\ell,k,t}^{s}=1$ indicates that batch $k$ on line $\ell$ is processed by unit $s$ at time $t$, while $x_{\ell,k,t}^{s}=0$ indicates that the batch is not processed by this unit and its SoW remains unchanged. $P_{\ell,k,t}^{s}$ is the actual processing power, $\Delta t^{\mathrm{PL}}$ is the scheduling interval of PLs, and $E_{\ell,k}^{s}$ is the total energy required to complete the corresponding stage.

\subsubsection{Process loads operation constraints}~\label{PLcons}

In short-process steelmaking, each batch sequentially passes through EAF, LF, and CC. To ensure process-feasible scheduling, the operation of PLs must satisfy the device power limits, device occupancy limits, non-interruptible processing requirements, stage sequence constraints, inter-stage transfer and waiting time limits, processing-duration limits, and daily production requirement. Let $\mathcal{A}=\{(\mathrm{EAF},\mathrm{LF}),(\mathrm{LF},\mathrm{CC})\}$ denote the set of adjacent process-stage pairs. For batch $k$ on production line $\ell$, $T_{\ell,k}^{s,\mathrm{st}}$ and $T_{\ell,k}^{s,\mathrm{ed}}$ denote the start and end time slots on unit $s$, respectively. The operation constraints of PLs are formulated as follows.

\textbf{Constraints}: $\forall \ell\in\mathcal{L}, \forall k\in\mathcal{K}_{\ell}, \forall s\in\mathcal{S}, \forall t\in\mathcal{T}$
\begin{subequations}
\label{eq:PL_constraints}
\begin{align}
&\underline{P}^{s}x_{\ell,k,t}^{s}
\leq
P_{\ell,k,t}^{s}
\leq
\overline{P}^{s}x_{\ell,k,t}^{s},
\label{eq:PL_power_bound} \\
&\sum_{k\in\mathcal{K}_{\ell}}
x_{\ell,k,t}^{s}
\leq
1,
\label{eq:PL_device_occupancy} \\
&t
\geq
T_{\ell,k}^{s,\mathrm{st}}
-M\left(1-x_{\ell,k,t}^{s}\right),
\label{eq:PL_start_time} \\
&t
\leq
T_{\ell,k}^{s,\mathrm{ed}}+M\left(1-x_{\ell,k,t}^{s}\right),
\label{eq:PL_end_time} \\
&\sum_{t\in\mathcal{T}}
x_{\ell,k,t}^{s}=T_{\ell,k}^{s,\mathrm{ed}}-T_{\ell,k}^{s,\mathrm{st}}+1,
\label{eq:PL_continuity} \\
&\mathrm{SoW}_{\ell,k,t}^{s}
\geq x_{\ell,k,t}^{s}
-x_{\ell,k,t+1}^{s},
\quad
\forall t\in\mathcal{T}\setminus\{|\mathcal{T}|\},
\label{eq:PL_completion_shutdown} \\
&\underline{D}^{s}
\leq
\left(
T_{\ell,k}^{s,\mathrm{ed}}
-T_{\ell,k}^{s,\mathrm{st}}+1
\right)
\Delta t^{\mathrm{PL}}
\leq
\overline{D}^{s},
\label{eq:PL_duration} \\
&\underline{W}^{s,r}
\leq
\left(
T_{\ell,k}^{r,\mathrm{st}}-T_{\ell,k}^{s,\mathrm{ed}}-1
\right)
\Delta t^{\mathrm{PL}}
\leq
\overline{W}^{s,r},
\quad
\forall (s,r)\in\mathcal{A},
\label{eq:PL_transfer_waiting} \\
&T_{\ell,k+1}^{s,\mathrm{st}}
\geq
T_{\ell,k}^{s,\mathrm{st}},
\quad
\forall k\in\mathcal{K}_{\ell}\setminus\{|\mathcal{K}_{\ell}|\},
\label{eq:PL_batch_order} \\
&\left(
T_{\ell,k+1}^{\mathrm{EAF},\mathrm{st}}-T_{\ell,k}^{\mathrm{EAF},\mathrm{ed}}-1
\right)
\Delta t^{\mathrm{PL}}
\geq
D^{\mathrm{EAF,gap}},
\quad
\forall k\in\mathcal{K}_{\ell}\setminus\{|\mathcal{K}_{\ell}|\},
\label{eq:PL_EAF_gap} \\
&\sum_{\ell\in\mathcal{L}}
\sum_{k\in\mathcal{K}_{\ell}}
\mathrm{SoW}_{\ell,k,|\mathcal{T}|}^{\mathrm{CC}}
\geq
N^{\mathrm{tar}},
\label{eq:PL_quota} \\
&x_{\ell,k,t}^{s}\in\{0,1\},
\quad
T_{\ell,k}^{s,\mathrm{st}},T_{\ell,k}^{s,\mathrm{ed}}\in\mathbb{Z}_{+}.
\label{eq:PL_variable_domain}
\end{align}
\end{subequations}

\noindent where $\mathcal{L}$, $\mathcal{K}_{\ell}$, $\mathcal{S}$, and $\mathcal{T}$ denote the sets of production lines, batches on line $\ell$, process units, and scheduling intervals, respectively, with $\mathcal{S}=\{\mathrm{EAF},\mathrm{LF},\mathrm{CC}\}$. $P_{\ell,k,t}^{s}$ is the actual processing power of batch $k$ on line $\ell$ assigned to unit $s$ at time $t$, and $x_{\ell,k,t}^{s}$ is the binary processing-status variable. Specifically, $x_{\ell,k,t}^{s}=1$ indicates that the batch is processed by unit $s$ at time $t$, while $x_{\ell,k,t}^{s}=0$ indicates otherwise. $T_{\ell,k}^{s,\mathrm{st}}$ and $T_{\ell,k}^{s,\mathrm{ed}}$ are the start and end time slots of batch $k$ on unit $s$, respectively. $\underline{P}^{s}$ and $\overline{P}^{s}$ are the minimum and maximum operating powers of unit $s$. $\underline{D}^{s}$ and $\overline{D}^{s}$ denote the minimum and maximum processing durations on unit $s$. $\underline{W}^{s,r}$ and $\overline{W}^{s,r}$ denote the minimum transfer time and maximum waiting time between adjacent stages $(s,r)$. $D^{\mathrm{EAF,gap}}$ represents the minimum time gap between two consecutive EAF batches on the same line, and $N^{\mathrm{tar}}$ is the target number of completed batches within the scheduling horizon.

Constraint \eqref{eq:PL_power_bound} imposes the operating power limits of each process unit, ensuring that the processing power is zero when the corresponding batch is not processed and remains within the admissible range when it is processed. Constraint \eqref{eq:PL_device_occupancy} ensures that each unit on each production line processes at most one batch at any time. Constraints \eqref{eq:PL_start_time}--\eqref{eq:PL_continuity} define a continuous processing interval for each batch on each unit, thereby enforcing non-interruptible processing. Constraint \eqref{eq:PL_completion_shutdown} ensures that a batch can leave the current unit only after the corresponding stage is completed. Constraint \eqref{eq:PL_duration} bounds the processing duration of each batch on each unit. Constraint \eqref{eq:PL_transfer_waiting} enforces the process coupling between EAF--LF and LF--CC by imposing both the minimum transfer time and maximum waiting time. Constraint \eqref{eq:PL_batch_order} preserves the batch order within each production line. Constraint \eqref{eq:PL_EAF_gap} imposes the minimum time gap between consecutive EAF batches. Constraint \eqref{eq:PL_quota} imposes the daily production quota requirement.

\section{Sequential heterogeneous-agent coordination framework}~\label{methodology}

\subsection{Reference-free coordination architecture}

\begin{figure}[pos=htbp] 
  \footnotesize\rmfamily   \setlength{\abovecaptionskip}{-0.1cm}  
    \setlength{\belowcaptionskip}{-0.1cm} 
  \begin{center}  \includegraphics[width=1\columnwidth]{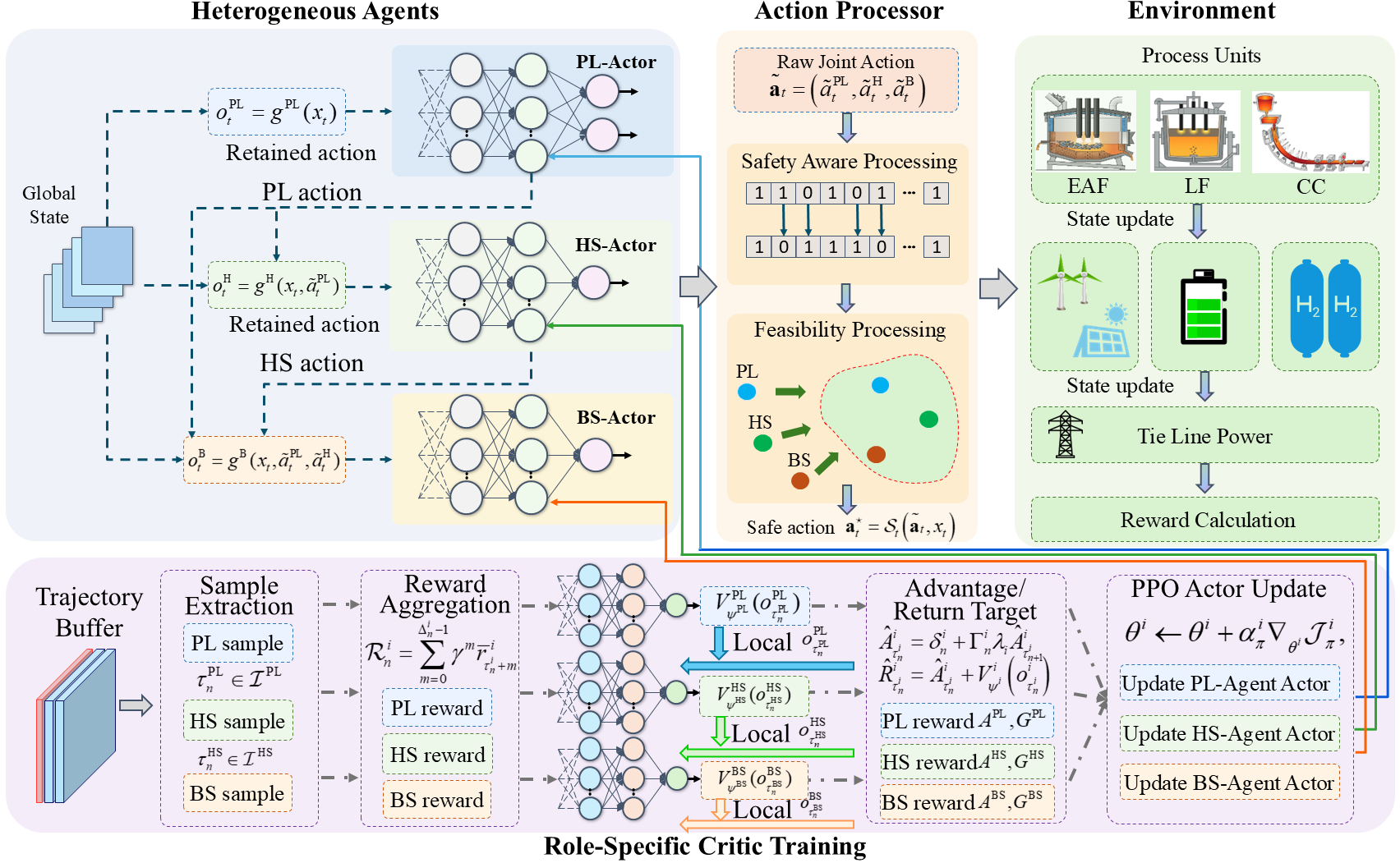}
     \caption{\rmfamily Overall workflow of the proposed SHAC framework.}
     \captionsetup{justification=centering}\label{Flowchart}
  \end{center}
  \vspace{-2.5em}
\end{figure}

To shape a grid-friendly TLP trajectory, this paper proposes the sequential heterogeneous-agent coordination (SHAC) framework for reference-free and safety-aware operation of heterogeneous agents. Short-process steelmaking PLs are governed by batch operations and strong inter-stage coupling. Although their scheduling flexibility can redistribute electricity demand over relatively long intervals, discrete production transitions still induce pronounced power variations. To address this issue, the system is decomposed into three functionally heterogeneous agents: a PL agent, a HS agent, and a BS agent. The PL agent adjusts production schedules within process-feasible boundaries to provide low-frequency coarse load shaping. The HS agent performs slow-timescale energy buffering to regulate the medium-timescale TLP residual, while the BS agent provides fast residual regulation to suppress minute-level TLP fluctuations. Considering the distinct response speeds and operating mechanisms of PLs, hydrogen storage, and battery storage, a multi-timescale asynchronous decision structure is adopted under a role-aware training and decentralized execution paradigm. During training, role-specific critics evaluate the corresponding agents according to their coordination objectives. During execution, the PL, HS, and BS agents make decisions based on cross-role local observations at 5-min, 15-min, and 1-min intervals, respectively, while safety-aware action processing ensures the feasibility of production logic and storage physical constraints. The overall workflow of SHAC is illustrated in Fig.~\ref{Flowchart}.

\subsection{Heterogeneous Markov game formulation}

The real-time dispatch problem is formulated as a cooperative heterogeneous Markov game with asynchronous decision epochs. The basic execution interval is set to $1$-min. The PL, HS, and BS agents make decisions at $5$-min, $15$-min, and $1$-min intervals, respectively. Between two consecutive decision epochs, the latest valid PL or HS action is retained for execution. The heterogeneous Markov game is defined as:
\begin{equation}
\mathcal{G}
=\left\langle
\mathcal{N},
\mathcal{X},
\{\mathcal{O}^{i}\}_{i\in\mathcal{N}},
\{\mathcal{A}^{i}\}_{i\in\mathcal{N}},
\{\mathcal{I}^{i}\}_{i\in\mathcal{N}},
\mathcal{P},
\{r^{i}\}_{i\in\mathcal{N}},
\gamma
\right\rangle ,
\end{equation}

\noindent where $\mathcal{N}=\{\mathrm{PL},\mathrm{HS},\mathrm{BS}\}$ denotes the set of heterogeneous agents. $\mathcal{X}$ is the global state space, $\mathcal{O}^{i}$ and $\mathcal{A}^{i}$ are the observation and action spaces of agent $i$, $\mathcal{I}^{i}$ is its valid decision epoch set, $\mathcal{P}$ is the transition kernel, $r^{i}$ is the role-specific reward, and $\gamma\in[0,1)$ is the discount factor. Let $\mathcal{T}^{\mathrm{M}}=\{0,1,\ldots,T^{\mathrm{M}}-1\}$ denote the set of minute-level operation periods. The valid decision epochs are $\mathcal{I}^{\mathrm{PL}}=\{t\in\mathcal{T}^{\mathrm{M}}\mid t\bmod 5=0\}$, $\mathcal{I}^{\mathrm{H}}=\{t\in\mathcal{T}^{\mathrm{M}}\mid t\bmod 15=0\}$, and $\mathcal{I}^{\mathrm{B}}=\mathcal{T}^{\mathrm{M}}$. The action-retention mapping is defined as:

\begin{equation}
\kappa_m(t)
=m\left\lfloor\frac{t}{m}\right\rfloor,
\end{equation}

\noindent then the PL scheduling action is retained over $[\kappa_5(t),\kappa_5(t)+4]$, the HS action is retained over $[\kappa_{15}(t),\kappa_{15}(t)+14]$, and the BS action is updated at each minute-level period. To preserve the Markov property under asynchronous action retention, the global state $x_t\in\mathcal{X}$ includes operating states, exogenous forecasts, retained actions, and temporal phase indicators:
\begin{equation}
x_t=\Big[\Xi_t^{\mathrm{PL}},
E_t^{\mathrm{B}},
E_t^{\mathrm{H}},
P_{t-1}^{\mathrm{TL}},
\bar a_t^{\mathrm{PL}},
\bar \rho_t^{\mathrm{H}},
\{\widehat P_{\tau|t}^{\mathrm{RES}},\widehat c_{\tau|t}^{\mathrm{e}}\}_{\tau=t}^{t+H_s-1},
\boldsymbol{\chi}_t
\Big],
\end{equation}

\noindent where $\Xi_t^{\mathrm{PL}}$ collects the operating states of EAFs, LFs, and CCs, including batch progress, remaining processing time, inter-stage waiting status, admissible startup candidates, and remaining production tasks. $\bar a_t^{\mathrm{PL}}=a_{\kappa_5(t)}^{\mathrm{PL}}$ and $\bar \rho_t^{\mathrm{H}}=\rho_{\kappa_{15}(t)}^{\mathrm{H}}$ denote the retained PL scheduling action and the retained normalized HS action, respectively. $\boldsymbol{\chi}_t$ collects quota-related, time-related, TLP-related, storage-related, process-urgency, and temporal phase indicators.

During decentralized execution, SHAC adopts sequentially conditioned observations. The PL agent acts first based on the current production state and previously available storage-support information. The HS agent then observes the current PL decision, while the BS agent further observes the realized PL power and executable HS action. The information flow is represented as:
\begin{subequations}
\label{eq:sequential_observation_maps}
\begin{align}
&o_t^{\mathrm{PL}}
=g^{\mathrm{PL}}(x_t),
\label{eq:pl_observation_map}\\
&o_t^{\mathrm{H}}=g^{\mathrm{H}}\left(x_t,\widetilde a_t^{\mathrm{PL}}\right),
\label{eq:hs_observation_map}\\
&o_t^{\mathrm{B}}=
g^{\mathrm{B}}\left(x_t,\widetilde a_t^{\mathrm{PL}},\widetilde a_t^{\mathrm{H}}\right),
\label{eq:bs_observation_map}
\end{align}
\end{subequations}

\noindent where $g_i(\cdot)$ is a deterministic observation-construction map. The corresponding observations are represented as:
\begin{subequations}
\label{eq:agent_observations}
\begin{align}
&o_t^{\mathrm{PL}}
=\Big[\Xi_t^{\mathrm{PL}},
\boldsymbol{\zeta}_{t-1}^{\mathrm{B}},
\boldsymbol{\zeta}_{t-1}^{\mathrm{H}},
P_{t-1}^{\mathrm{TL}},
\bar \rho_{t-1}^{\mathrm{H}},
\{\widehat P_{\tau|t}^{\mathrm{RES}},\widehat c_{\tau|t}^{\mathrm{e}}\}_{\tau=t}^{t+H_s-1},
\boldsymbol{\chi}_t^{\mathrm{PL}}
\Big],
\label{eq:pl_observation}\\
&o_t^{\mathrm{H}}
=\Big[E_t^{\mathrm{H}},
E_{t-1}^{\mathrm{B}},
\widetilde a_t^{\mathrm{PL}},
\{\widehat P_{\tau|t}^{\mathrm{PL}}\}_{\tau=t}^{t+H_s-1},
P_{t-1}^{\mathrm{TL}},
\bar \rho_t^{\mathrm{H}},
\{\widehat P_{\tau|t}^{\mathrm{RES}},\widehat c_{\tau|t}^{\mathrm{e}}\}_{\tau=t}^{t+H_s-1},
\boldsymbol{\chi}_t^{\mathrm{H}}
\Big],
\label{eq:hs_observation}\\
&o_t^{\mathrm{B}}
=\Big[E_t^{\mathrm{B}},
E_t^{\mathrm{H}},
P_t^{\mathrm{PL}},
u_t^{\mathrm{H},\star},
P_{t-1}^{\mathrm{TL}},
\{\widehat P_{\tau|t}^{\mathrm{RES}},\widehat c_{\tau|t}^{\mathrm{e}}\}_{\tau=t}^{t+H_s-1},
\boldsymbol{\chi}_t^{\mathrm{B}}
\Big],
\label{eq:bs_observation}
\end{align}
\end{subequations}

\noindent where $\boldsymbol{\zeta}_{t-1}^{\mathrm{B}}$ and $\boldsymbol{\zeta}_{t-1}^{\mathrm{H}}$ denote the previously available storage support indicators provided to the PL agent. $\{\widehat P_{\tau|t}^{\mathrm{PL}}\}_{\tau=t}^{t+H_s-1}$ denotes the PL power inferred from the current process state and retained production mode. These cross-role observations enable the PL agent to account for storage flexibility, the HS agent to anticipate 5-min scale TLP residuals induced by PL scheduling, and the BS agent to regulate minute-level TLP fluctuations after the HS action.

To reduce the discrete action dimension, the PL agent makes scheduling decisions only for active-frontier batches. For device $s\in\mathcal{S}$ on production line $\ell$, the PL action is defined over the active-frontier set $\mathcal{F}_{\ell,s,\tau}$ as follows:
\begin{subequations}
\label{eq:agent_actions}
\begin{align}
&\begin{aligned}[t]
&a_{\tau}^{\mathrm{PL}}=
\left(a_{\tau}^{x,\mathrm{PL}},
\boldsymbol{\rho}_{\tau}^{\mathrm{PL}}\right)=\left(
\left\{a_{\ell,s,k,\tau}^{x,\mathrm{PL}}\right\},
\left\{\rho_{\ell,s,k,\tau}^{\mathrm{PL}}\right\}
\right),\quad \widetilde a_t^{\mathrm{PL}}=a_{\kappa_5(t)}^{\mathrm{PL}},\\
&a_{\ell,s,k,\tau}^{x,\mathrm{PL}}\in\{0,1\},
\quad \rho_{\ell,s,k,\tau}^{\mathrm{PL}}\in[0,1],
\quad \tau\in\mathcal{I}^{\mathrm{PL}},
\ell\in\mathcal{L}, s\in\mathcal{S}, k\in\mathcal{F}_{\ell,s,\tau},
\end{aligned}
\label{eq:pl_action}\\
&\widetilde a_t^{\mathrm{H}}=\rho_{\kappa_{15}(t)}^{\mathrm{H}}, \quad
\rho_{\kappa_{15}(t)}^{\mathrm{H}}\in[-1,1],
\label{eq:hydrogen_action}\\
&\widetilde a_t^{\mathrm{B}}=\rho_t^{\mathrm{B}}, \quad \rho_t^{\mathrm{B}}\in[-1,1].
\label{eq:battery_action}
\end{align}
\end{subequations}

\noindent where $a_{\ell,s,k,\tau}^{x,\mathrm{PL}}$ denotes the connection intention between device $s$ and active-frontier batch $k$, and $\rho_{\ell,s,k,\tau}^{\mathrm{PL}}$ denotes the normalized processing-power command associated with this candidate connection. $\rho_t^{\mathrm{B}}$ and $\rho_{\kappa_{15}(t)}^{\mathrm{H}}$ denote the normalized signed actions of BS and HS, respectively. At valid decision epochs, the raw actions are sampled from non-parameter-sharing heterogeneous policies:
\begin{subequations}
\label{eq:action_sampling}
\begin{align}
&a_{\tau}^{\mathrm{PL}}
\sim
\pi_{\theta^{\mathrm{PL}}}^{\mathrm{PL}}
\left(\cdot\mid o_{\tau}^{\mathrm{PL}}\right),
\quad
\tau\in\mathcal{I}^{\mathrm{PL}},
\label{eq:pl_action_sampling}\\
&\rho_{\tau}^{\mathrm{H}}
\sim
\pi_{\theta^{\mathrm{H}}}^{\mathrm{H}}
\left(\cdot\mid o_{\tau}^{\mathrm{H}}\right),
\quad
\tau\in\mathcal{I}^{\mathrm{H}},
\label{eq:hydrogen_action_sampling}\\
&\rho_t^{\mathrm{B}}
\sim
\pi_{\theta^{\mathrm{B}}}^{\mathrm{B}}
\left(\cdot\mid o_t^{\mathrm{B}}\right),
\quad
t\in\mathcal{I}^{\mathrm{B}}.
\label{eq:battery_action_sampling}
\end{align}
\end{subequations}

The minute-level execution distribution of SHAC is induced by non-parameter-sharing local policies and factorized according to the sequential information flow:
\begin{subequations}
\label{eq:joint_policy}
\begin{align}
&\boldsymbol{\pi}_{\theta}
\left(\widetilde{\mathbf a}_t\mid x_t\right)=
\pi_{\theta^{\mathrm{PL}}}^{\mathrm{PL}}\left(
\widetilde a_t^{\mathrm{PL}}\mid o_t^{\mathrm{PL}}\right)
\pi_{\theta^{\mathrm{H}}}^{\mathrm{H}}\left(
\widetilde a_t^{\mathrm{H}}\mid o_t^{\mathrm{H}}\left(\widetilde a_t^{\mathrm{PL}}\right)\right)
\pi_{\theta^{\mathrm{B}}}^{\mathrm{B}}\left(
\widetilde a_t^{\mathrm{B}}\mid o_t^{\mathrm{B}}\left(\widetilde a_t^{\mathrm{PL}},\widetilde a_t^{\mathrm{H}}\right)\right),
\label{eq:joint_policy_factorization}\\
&\widetilde{\mathbf a}_t
=\left(
\widetilde a_t^{\mathrm{PL}},
\widetilde a_t^{\mathrm{H}},
\widetilde a_t^{\mathrm{B}}\right),\quad\theta=
\{\theta^{\mathrm{PL}},\theta^{\mathrm{H}},\theta^{\mathrm{B}}\},
\label{eq:joint_policy_variables}
\end{align}
\end{subequations}

\noindent where $\boldsymbol{\pi}_{\theta}$ denotes the joint execution distribution induced by non-parameter-sharing local policies. The factorization follows the SHAC coordination order, i.e., PL $\rightarrow$ HS $\rightarrow$ BS, where HS conditions its action on the current PL decision and BS further conditions its action on the PL and HS actions. The policy parameters are updated through role-specific training objectives. The normalized raw joint action $\widetilde{\mathbf a}_t$ is mapped to an executable physical action through the safety-aware action processor:
\begin{equation}
\mathbf a_t^\star
=\mathcal{S}_t\left(
\widetilde{\mathbf a}_t,x_t\right),
\end{equation}

\noindent where $\mathcal{S}_t(\cdot)$ denotes the safety-processing mapping induced by production logic, storage physical constraints, and TLP-related operational requirements. Under $\mathbf a_t^\star$, the minute-level system transition is described by
\begin{subequations}
\label{eq:system_transition}
\begin{align}
&P_t^{\mathrm{PL}}
=\sum_{\ell\in\mathcal{L}}\sum_{s\in\mathcal{S}}
\sum_{k\in\mathcal{K}_{\ell}}P_{\ell,k,t}^{s,\star},
\label{eq:pl_power_transition}\\
&P_t^{\mathrm{TL}}=P_t^{\mathrm{PL}}+u_t^{\mathrm{H},\star}+u_t^{\mathrm{B},\star}
-P_t^{\mathrm{RES}},
\label{eq:tlp_transition}\\
&x_{t+1}\sim\mathcal{P}
\left(\cdot\mid x_t,\mathbf a_t^\star\right),
\label{eq:state_transition}
\end{align}
\end{subequations}

\noindent where $P_{\ell,k,t}^{s,\star}$ is the executable processing power, and $u_t^{\mathrm{H},\star}$ and $u_t^{\mathrm{B},\star}$ are the executable signed storage powers of HS and BS, respectively. The cooperative objective of SHAC is defined by the role-specific returns:
\begin{equation}
\max_{\theta} \enspace J_{\mathrm{SHAC}}(\theta)
=\mathbb{E}_{\boldsymbol{\pi}_{\theta},\mathcal{P}}
\left[
\sum_{i\in\mathcal{N}}
\sum_{t\in\mathcal{I}^{i}}
\gamma^{t}r_t^{i}
\right].
\end{equation}

\subsection{Safety-aware action processing}

Raw outputs of the heterogeneous local policies may violate PL operating logic or storage physical constraints. Therefore, a safety-aware action processor is introduced to map normalized policy outputs into executable physical actions before environment transition. The processor consists of process-knowledge-based masking for PL discrete actions and feasibility projection for PL, HS, and BS continuous power actions. The overall mapping is expressed as:

\begin{subequations}
\label{eq:safety_action_processing}
\begin{align}
&\mathbf a_t^\star=\mathcal{S}_t
\left(\widetilde{\mathbf a}_t,x_t\right)=\left(
a_t^{x,\mathrm{PL},\star}, \mathbf P_t^{\mathrm{PL},\star},
u_t^{\mathrm{H},\star}, u_t^{\mathrm{B},\star}\right),
\label{eq:safety_mapping}\\
&m_{\tau}^{\mathrm{PL}}\left(a_{\tau}^{x,\mathrm{PL}}\right)=
\begin{cases}
1, & a_{\tau}^{x,\mathrm{PL}}
\in
\mathcal{A}_{\tau}^{x,\mathrm{PL,safe}}(x_{\tau}),\\
0, & \mathrm{otherwise},
\end{cases}\quad\tau\in\mathcal{I}^{\mathrm{PL}},
\label{eq:pl_mask}\\
&\widetilde{\pi}_{\theta^{\mathrm{PL}}}^{x,\mathrm{PL}}
\left(a_{\tau}^{x,\mathrm{PL}}\mid o_{\tau}^{\mathrm{PL}}
\right)=\frac{
\pi_{\theta^{\mathrm{PL}}}^{x,\mathrm{PL}}\left(
a_{\tau}^{x,\mathrm{PL}}\mid o_{\tau}^{\mathrm{PL}}\right)
m_{\tau}^{\mathrm{PL}}\left(
a_{\tau}^{x,\mathrm{PL}}\right)}{
\sum\limits_{a'\in\mathcal{A}_{\tau}^{x,\mathrm{PL}}}
\pi_{\theta^{\mathrm{PL}}}^{x,\mathrm{PL}}\left(
a'\mid o_{\tau}^{\mathrm{PL}}\right)
m_{\tau}^{\mathrm{PL}}\left(a'\right)},\quad
a_{\tau}^{x,\mathrm{PL},\star}
\sim
\widetilde{\pi}_{\theta^{\mathrm{PL}}}^{x,\mathrm{PL}}
\left(\cdot\mid o_{\tau}^{\mathrm{PL}}\right),
\label{eq:masked_policy}\\
&P_{\ell,s,k,t}^{\mathrm{PL},\star}=a_{\ell,s,k,\kappa_5(t)}^{x,\mathrm{PL},\star}
\left[\underline{P}_{\ell,s,k,t}^{\mathrm{PL}}+
\rho_{\ell,s,k,\kappa_5(t)}^{\mathrm{PL}}\left(
\overline{P}_{\ell,s,k,t}^{\mathrm{PL}}-
\underline{P}_{\ell,s,k,t}^{\mathrm{PL}}\right)\right],
\label{eq:pl_power_mapping}\\
&\rho_t^{\mathrm{H}}
=\rho_{\kappa_{15}(t)}^{\mathrm{H}},\quad u_t^{q,\mathrm{raw}}=
\rho_t^q\overline{P}^{q},\quad
q\in\{\mathrm{H},\mathrm{B}\},
\label{eq:storage_raw_power}\\
&u_t^{q,\star}=\min\left\{
\overline{u}_t^{q},\max\left\{
u_t^{q,\mathrm{raw}},\underline{u}_t^{q}\right\}\right\},
\quad q\in\{\mathrm{H},\mathrm{B}\},
\label{eq:storage_power_projection}
\end{align}
\end{subequations}

\noindent where the safe PL action set $\mathcal{A}_{\tau}^{x,\mathrm{PL,safe}}(x_{\tau})$ is constructed from device occupancy, EAF--LF--CC process sequence, non-interruptible processing, minimum transfer time, maximum waiting time, EAF inter-batch interval, and batch-order constraints. Mandatory continuation and downstream transfer actions are embedded when required by process continuity or waiting-time limits. Equation \eqref{eq:masked_policy} renormalizes the discrete PL policy over the safe action set before execution, while \eqref{eq:pl_power_mapping} maps the normalized PL power command to the feasible processing power interval of the selected batch-device connection. $u_t^{q,\mathrm{raw}}$ denotes the raw signed power command converted from the normalized action, where positive and negative values correspond to charging and discharging, respectively. The bounds $\underline{u}_t^{q}$ and $\overline{u}_t^{q}$ denote the lower and upper executable signed power limits determined by the storage power rating, energy state, conversion efficiency, and retained action duration. Equation \eqref{eq:storage_power_projection} projects the raw storage command onto the feasible physical interval before execution. The resulting executable actions are then evaluated by the role-specific rewards in the subsequent training process.

\subsection{Role-specific reward design}

Following the grid-friendly TLP characterization in Section~\ref{Grid-friendly_TLP_characterization}, SHAC adopts role-specific rewards to align the learning signals with the physical responsibilities of heterogeneous agents, thereby mitigating ambiguous credit assignment and improving training stability. The PL reward is designed for production profitability, grid-stress-responsive electricity consumption, contract demand management, and process feasibility. The HS reward focuses on medium-timescale TLP regulation and long-duration storage sustainability, while the BS reward focuses on minute-level TLP smoothing and fast regulation sustainability. The role-specific rewards are formulated as:
\begin{subequations}
\label{eq:role_specific_rewards}
\begin{align}
&r_t^{\mathrm{PL}}=R_t^{\mathrm{fin}}-C_t^{\mathrm{grd}}-
C_t^{\mathrm{loss}}+\mathbb{I}_{t=T^{\mathrm{M}}-1}R_T^{\mathrm{PL}},
\quad t\in\mathcal{I}^{\mathrm{PL}},
\label{eq:pl_reward}\\
&r_t^{\mathrm{H}}=-C_t^{\mathrm{H,res}}-C_t^{\mathrm{H,sto}}+
\mathbb{I}_{t=T^{\mathrm{M}}-1}
\left(R_T^{\mathrm{H}}+R_T^{\mathrm{ex}}\right),\quad t\in\mathcal{I}^{\mathrm{H}},
\label{eq:hs_reward}\\
&r_t^{\mathrm{B}}=-C_t^{\mathrm{ramp}}-C_t^{\mathrm{B,sto}}+
\mathbb{I}_{t=T^{\mathrm{M}}-1}R_T^{\mathrm{B}},\quad
t\in\mathcal{I}^{\mathrm{B}},
\label{eq:bs_reward}
\end{align}
\end{subequations}

\noindent where HS represents long-duration storage, no intra-day terminal SOC constraint is imposed. In contrast, BS is a short-duration storage resource for fast regulation, and its terminal SOC penalty is retained to preserve the regulation capability for subsequent operation periods. The reward components in \eqref{eq:role_specific_rewards} are defined as:
\begin{subequations}
\label{eq:reward_components}
\begin{align}
&C_t^{\mathrm{grd}}=c_t^{\mathrm{e}}\left[P_t^{\mathrm{TL}}\right]^+\Delta t
+\left(\mu^{\mathrm{dem}}-1\right)c_t^{\mathrm{e}}\left[
\left[P_t^{\mathrm{TL}}\right]^+-D^{\mathrm{dem}}\right]^+\Delta t,
\label{eq:grid_cost}\\
&C_t^{\mathrm{ramp}}=c^{\mathrm{ramp}}\left|\left|\Delta P_t^{\mathrm{TL}}\right|
-R^{\max}\right|,\label{eq:tlp_ramp_penalty}\\
&C_t^{q,\mathrm{sto}}=c^{q}\left|u_t^{q,\star}\right|
\Delta t,\quad q\in\{\mathrm{B},\mathrm{H}\},\label{eq:storage_throughput_cost}\\
&C_t^{\mathrm{H,res}}=c^{\mathrm{H,res}}
\left(\overline{P}_{t}^{\mathrm{TL,H}}\right)^2,
\quad \overline{P}_{t}^{\mathrm{TL,H}}=\frac{1}{H_{\mathrm{H}}}
\sum_{\tau=t}^{t+H_{\mathrm{H}}-1}
P_{\tau}^{\mathrm{TL}},
\label{eq:hs_residual_penalty}\\
&C_t^{\mathrm{loss}}=c^{\mathrm{hm}}n_t^{\mathrm{hm}}+
c^{\mathrm{sp}}n_t^{\mathrm{sp}},
\label{eq:process_loss_cost}\\
&R_t^{\mathrm{fin}}=\rho^{\mathrm{h}}\Delta N_t^{\mathrm{fin}},
\quad \Delta N_t^{\mathrm{fin}}=N_t^{\mathrm{fin}}-N_{t-1}^{\mathrm{fin}},
\label{eq:completion_revenue}\\
&R_T^{\mathrm{PL}}=-c^{\mathrm{ms}}(N^{\mathrm{tar}}-N_T^{\mathrm{fin}}),
\label{eq:pl_terminal_reward}\\
&R_T^{\mathrm{B}}=-c^{\mathrm{soc}}
\left|\frac{E_T^{\mathrm{B}}}{\overline{E}^{\mathrm{B}}}-\xi^{\mathrm{B,tar}}\right|,
\label{eq:battery_terminal_reward}
\end{align}
\end{subequations}

\noindent where $C_t^{\mathrm{grd}}$ denotes the electricity purchase and contract demand exceedance cost, $c_t^{\mathrm{e}}$ is the electricity price. $C_t^{\mathrm{ramp}}$ measures the deviation between the minute-level TLP variation $\Delta P_t^{\mathrm{TL}}$ and the maximum acceptable ramping rate $R^{\max}$. $C_t^{q,\mathrm{sto}}$ denotes the operating cost of storage. $C_t^{\mathrm{H,res}}$ penalizes the medium-timescale TLP residual over the HS regulation horizon $H_{\mathrm{H}}$. $C_t^{\mathrm{loss}}$ and $R_t^{\mathrm{fin}}$ characterize process losses and production completion, respectively. $R_T^{\mathrm{PL}}$ penalizes terminal production shortfall, while $R_T^{\mathrm{B}}$ penalizes terminal BS SOC deviation.

\subsection{Role-aware multi-timescale training and decentralized execution}

To train the heterogeneous PL, HS, and BS agents with different decision intervals and action spaces, SHAC adopts a role-aware multi-timescale actor--critic training scheme. The common coordination objective defined in the heterogeneous Markov game is optimized through non-parameter-sharing local actors, role-specific critics, and clipped policy updates. Although the agents are evaluated by role-specific reward signals and value functions, their samples are collected from the same IMG environment under the sequential SHAC rollout. Therefore, the training process remains coordinated through the jointly induced TLP trajectory, cross-role observations, retained actions, and shared environment transition.

Given the behavior policy parameters $\theta_{\mathrm{old}}=\{\theta_{\mathrm{old}}^{i}\}_{i\in\mathcal{N}}$, the trajectory distribution used for policy updating is induced by the non-parameter-sharing local policies and the safety-aware transition process. Let $\mathcal{N}_t=\{i\in\mathcal{N}\mid t\in\mathcal{I}^i\}$ denote the ordered set of agents that make new decisions at time $t$, following the SHAC order PL $\rightarrow$ HS $\rightarrow$ BS when multiple agents act simultaneously. The trajectory distribution is expressed as:
\begin{equation}
\mathbb{P}_{\theta_{\mathrm{old}},\mathcal{P}}(\zeta)
=
\rho_0(x_0)
\prod_{i\in\mathcal{N}}
\prod_{\tau\in\mathcal{I}^{i}}
\pi_{\theta_{\mathrm{old}}^{i}}^{i}
\left(
a_{\tau}^{i}
\mid
o_{\tau}^{i}
\right)
\prod_{t=0}^{T^{\mathrm{M}}-1}
\mathcal{P}
\left(
x_{t+1}
\mid
x_t,
\mathcal{S}_t
\left(
\widetilde{\mathbf a}_t,x_t
\right)
\right),
\label{eq:shac_trajectory_distribution}
\end{equation}

\noindent where $\zeta$ denotes the episode trajectory. The raw joint action at the minute-level execution step is composed of the latest valid actions retained from the corresponding agents: $\widetilde{\mathbf a}_t=\left(a_{\kappa_5(t)}^{\mathrm{PL}},
\rho_{\kappa_{15}(t)}^{\mathrm{HS}}, \rho_t^{\mathrm{BS}}\right)$. The policy likelihoods in \eqref{eq:shac_trajectory_distribution} are evaluated only at the valid decision epochs of the corresponding agents, whereas the environment transition is evaluated at each minute-level period using the safety-processed action. This formulation avoids repeatedly counting the likelihoods of retained PL and HS actions within their holding windows. For policy updating, the log-likelihood of each agent at its valid decision epoch is denoted by:
\begin{subequations}
\label{eq:role_log_likelihood}
\begin{align}
&\ell_{\tau}^{\mathrm{PL}}
\left(
\theta^{\mathrm{PL}}
\right)
=
\log
\widetilde{\pi}_{\theta^{\mathrm{PL}}}^{x,\mathrm{PL}}
\left(
a_{\tau}^{x,\mathrm{PL}}
\mid
o_{\tau}^{\mathrm{PL}}
\right)
+
\log
\pi_{\theta^{\mathrm{PL}}}^{\rho,\mathrm{PL}}
\left(
\boldsymbol{\rho}_{\tau}^{\mathrm{PL}}
\mid
o_{\tau}^{\mathrm{PL}}
\right),
\quad
\tau\in\mathcal{I}^{\mathrm{PL}},
\label{eq:pl_log_likelihood}\\
&\ell_{\tau}^{\mathrm{HS}}
\left(
\theta^{\mathrm{HS}}
\right)
=
\log
\pi_{\theta^{\mathrm{HS}}}^{\mathrm{HS}}
\left(
\rho_{\tau}^{\mathrm{HS}}
\mid
o_{\tau}^{\mathrm{HS}}
\right),
\quad
\tau\in\mathcal{I}^{\mathrm{HS}},
\label{eq:hs_log_likelihood}\\
&\ell_{t}^{\mathrm{BS}}
\left(
\theta^{\mathrm{BS}}
\right)
=
\log
\pi_{\theta^{\mathrm{BS}}}^{\mathrm{BS}}
\left(
\rho_{t}^{\mathrm{BS}}
\mid
o_{t}^{\mathrm{BS}}
\right),
\quad
t\in\mathcal{I}^{\mathrm{BS}},
\label{eq:bs_log_likelihood}
\end{align}
\end{subequations}

\noindent where $\widetilde{\pi}_{\theta^{\mathrm{PL}}}^{x,\mathrm{PL}}$ denotes the masked discrete PL policy defined in \eqref{eq:masked_policy}. The PL log-likelihood is decomposed into the masked discrete-action likelihood and the continuous normalized processing-power likelihood, corresponding to the two branches generated from the same PL observation. All likelihoods are evaluated on the sampled policy actions before physical action mapping.

Each agent $i\in\mathcal{N}$ is assigned a role-specific critic $V_{\psi^i}^{i}(o_t^i)$. Let $\{\tau_n^i\}_{n=0}^{K_i}$ denote the augmented decision sequence of agent $i$, where $\tau_{K_i}^{i}=T^{\mathrm{M}}$ is the terminal epoch. The holding interval of the $n$th decision is defined as $\Delta_n^i=\tau_{n+1}^{i}-\tau_n^{i}$, $n=0,\ldots,K_i-1$. Since a retained PL or HS action affects multiple minute-level transitions, the reward assigned to a decision epoch is obtained by accumulating the corresponding minute-level reward components over its holding interval:
\begin{equation}
\mathcal{R}_{n}^{i}
=\sum_{m=0}^{\Delta_n^i-1}
\gamma^{m}
\bar r_{\tau_n^i+m}^{i},
\label{eq:role_interval_reward}
\end{equation}

\noindent where $\bar r_{t}^{i}$ denotes the minute-level contribution induced by the role-specific reward components defined in \eqref{eq:role_specific_rewards}. For the BS agent, whose action is updated every minute, $\mathcal{R}_{n}^{\mathrm{BS}}$ reduces to the minute-level reward. For the PL and HS agents, $\mathcal{R}_{n}^{i}$ aggregates the effect of a retained multi-minute decision.

Define $\Gamma_n^i=\gamma^{\Delta_n^i}$ as the discount factor across the holding interval of the $n$th valid decision of agent $i$. The temporal-difference error, generalized advantage estimate, return target, and critic loss are then defined as:
\begin{subequations}
\label{eq:role_specific_advantage}
\begin{align}
&\delta_n^i=\mathcal{R}_{n}^{i}+\Gamma_n^i
\bar V_{n+1}^{i}-V_{\psi^i}^{i}\left(o_{\tau_n^i}^{i}\right),
\label{eq:role_td_error}\\
&\widehat{A}_{\tau_n^i}^{i}=\delta_n^i+\Gamma_n^i\lambda_i
\widehat{A}_{\tau_{n+1}^i}^{i},\quad n=K_i-1,\ldots,0,
\label{eq:role_gae}\\
&\widehat{R}_{\tau_n^i}^{i}=\widehat{A}_{\tau_n^i}^{i}+V_{\psi^i}^{i}\left(
o_{\tau_n^i}^{i}\right),
\label{eq:role_return_target}\\
&\mathcal{L}_{V}^{i}\left(\psi^i\right)=
\mathbb{E}_{\zeta\sim\mathbb{P}_{\theta_{\mathrm{old}},\mathcal{P}},\,n=0,\ldots,K_i-1}\left[\left(V_{\psi^i}^{i}\left(o_{\tau_n^i}^{i}\right)
-\widehat{R}_{\tau_n^i}^{i}\right)^2\right],
\label{eq:role_critic_loss}
\end{align}
\end{subequations}

\noindent where $\bar V_{n+1}^{i}=V_{\psi^i}^{i}(o_{\tau_{n+1}^i}^{i})$ for nonterminal decision epochs and $\bar V_{n+1}^{i}=0$ at the terminal epoch. The terminal advantage is set as $\widehat{A}_{\tau_{K_i}^{i}}^{i}=0$. The advantage $\widehat{A}_{\tau_n^i}^{i}$ is computed along the valid decision sequence of agent $i$.

The PPO ratio is evaluated from the log-likelihood difference of the sampled policy action:
\begin{equation}
\varrho_{\tau}^{i}\left(\theta^i\right)=\exp\left[\ell_{\tau}^{i}
\left(\theta^i\right)-\ell_{\tau}^{i}\left(\theta_{\mathrm{old}}^i
\right)\right],\quad \tau\in\mathcal{I}^{i},\quad i\in\mathcal{N}.
\label{eq:role_policy_ratio}
\end{equation}

For the PL agent, the discrete component is computed under the masked policy. For the HS and BS agents, it is evaluated using the continuous signed-action likelihoods in \eqref{eq:hs_log_likelihood} and \eqref{eq:bs_log_likelihood}, respectively. Based on the role-specific advantage, the clipped actor objective is formulated as:
\begin{equation}
\mathcal{J}_{\pi}^{i}
\left(\theta^i\right)=
\mathbb{E}_{\zeta\sim\mathbb{P}_{\theta_{\mathrm{old}},\mathcal{P}},\,\tau\in\mathcal{I}^{i}}\left[\min\left(\varrho_{\tau}^{i}
\widehat{A}_{\tau}^{i},\operatorname{clip}
\left(\varrho_{\tau}^{i},1-\epsilon_{\pi},1+\epsilon_{\pi}\right)
\widehat{A}_{\tau}^{i}\right)\right],\quad i\in\mathcal{N}.
\label{eq:role_actor_objective}
\end{equation}

The parameters are updated by:
\begin{subequations}
\label{eq:parameter_update}
\begin{align}
&\theta^i
\leftarrow\theta^i+\alpha_{\pi}^{i}\nabla_{\theta^i}
\mathcal{J}_{\pi}^{i}\left(\theta^i\right),
\label{eq:actor_update}\\
&\psi^i\leftarrow\psi^i-\alpha_{V}^{i}\nabla_{\psi^i}
\mathcal{L}_{V}^{i}\left(\psi^i\right),\quad
i\in\{\mathrm{PL},\mathrm{HS},\mathrm{BS}\}.
\label{eq:critic_update}
\end{align}
\end{subequations}

The role-specific updates are coupled through the trajectory distribution in \eqref{eq:shac_trajectory_distribution}. The PL policy shapes the production-induced TLP profile, while the HS and BS policies are optimized under the trajectories sequentially affected by upstream decisions. In online execution, the critics are no longer required, and only the role-specific local actors are deployed for decentralized decision making. The actor outputs are then passed to the safety-aware action processor before being applied to the IMG environment.

\begin{algorithm}[htbp]
\caption{Sequential Heterogeneous-Agent Coordination Procedure}
\label{alg:SHAC}
\SetAlgoLined
\SetKwInOut{Input}{Input}
\SetKwInOut{Output}{Output}
\SetKwInOut{Result}{Result}

\Input{$\mathcal{D}$, $\mathcal{P}$, $\mathcal{S}_t(\cdot)$, actors $\{\pi_{\theta^i}^{i}\}$, critics $\{V_{\psi^i}^{i}\}$.}
\Output{Trained actors $\{\pi_{\theta^i}^{i}\}_{i\in\mathcal{N}}$.}

Set $\mathcal{N}$,
$\mathcal{I}^{\mathrm{PL}}$,
$\mathcal{I}^{\mathrm{HS}}$,
$\mathcal{I}^{\mathrm{BS}}$,
and $\kappa_m(t)$.\\[2pt]

\For{$e=1,2,\dots,E$}{
    Sample one day from $\mathcal{D}$ and initialize $x_0$.\\[2pt]

    \For{$t=0,1,\dots,T^{\mathrm{M}}-1$}{
        $o_t^{\mathrm{PL}}=g^{\mathrm{PL}}(x_t)$; sample $\widetilde a_t^{\mathrm{PL}}\sim\pi_{\theta^{\mathrm{PL}}}^{\mathrm{PL}}(\cdot|o_t^{\mathrm{PL}})$ if $t\in\mathcal{I}^{\mathrm{PL}}$, otherwise retain $\widetilde a_t^{\mathrm{PL}}=a_{\kappa_5(t)}^{\mathrm{PL}}$.\\[2pt]

        $o_t^{\mathrm{HS}}=g^{\mathrm{HS}}(x_t,\widetilde a_t^{\mathrm{PL}})$; sample $\widetilde a_t^{\mathrm{HS}}\sim\pi_{\theta^{\mathrm{HS}}}^{\mathrm{HS}}(\cdot|o_t^{\mathrm{HS}})$ if $t\in\mathcal{I}^{\mathrm{HS}}$, otherwise retain $\widetilde a_t^{\mathrm{HS}}=a_{\kappa_{15}(t)}^{\mathrm{HS}}$.\\[2pt]

        $o_t^{\mathrm{BS}}=g^{\mathrm{BS}}(x_t,\widetilde a_t^{\mathrm{PL}},\widetilde a_t^{\mathrm{HS}})$; sample $\widetilde a_t^{\mathrm{BS}}\sim\pi_{\theta^{\mathrm{BS}}}^{\mathrm{BS}}(\cdot|o_t^{\mathrm{BS}})$.\\[2pt]

        $\widetilde{\mathbf a}_t=(\widetilde a_t^{\mathrm{PL}},\widetilde a_t^{\mathrm{HS}},\widetilde a_t^{\mathrm{BS}})$,
        $\mathbf a_t^\star=\mathcal{S}_t(\widetilde{\mathbf a}_t,x_t)$ by \eqref{eq:safety_action_processing}.\\[2pt]

        Apply $\mathbf a_t^\star$, roll out $x_{t+1}$ by \eqref{eq:system_transition}, compute $\{r_t^i\}_{i\in\mathcal{N}}$ by \eqref{eq:role_specific_rewards}, and store valid samples at $t\in\mathcal{I}^{i}$.\\[2pt]
    }

    \For{$i\in\mathcal{N}$}{
        Compute $\mathcal{R}_n^i$, $\widehat A_{\tau_n^i}^{i}$, and $\widehat R_{\tau_n^i}^{i}$ by \eqref{eq:role_interval_reward}--\eqref{eq:role_return_target}.\\[2pt]
        Update $\psi^i$ and $\theta^i$ by \eqref{eq:critic_update} and \eqref{eq:actor_update}.\\[2pt]
    }
}

Discard $\{V_{\psi^i}^{i}\}_{i\in\mathcal{N}}$ and deploy $\{\pi_{\theta^i}^{i}\}_{i\in\mathcal{N}}$ with $\mathcal{S}_t(\cdot)$.\\[2pt]

\For{$t=0,1,\dots,T^{\mathrm{M}}-1$}{
    Generate $\widetilde{\mathbf a}_t$ sequentially by $\mathrm{PL}\rightarrow\mathrm{HS}\rightarrow\mathrm{BS}$, compute $\mathbf a_t^\star=\mathcal{S}_t(\widetilde{\mathbf a}_t,x_t)$, update $x_{t+1}$.\\[2pt]
}

\Result{$\{\mathbf a_t^\star,P_t^{\mathrm{TL}}\}_{t=0}^{T^{\mathrm{M}}-1}$.}
\end{algorithm}

\section{Case studies}~\label{Case_studies}

\subsection{Set-up}~\label{Set-up}

\subsubsection{Renewable Energy Sources and Market Data}

\begin{figure}[pos=htbp] 
  \footnotesize\rmfamily   \setlength{\abovecaptionskip}{-0.1cm}  
    \setlength{\belowcaptionskip}{-0.1cm} 
  \begin{center}  \includegraphics[width=0.9\columnwidth]{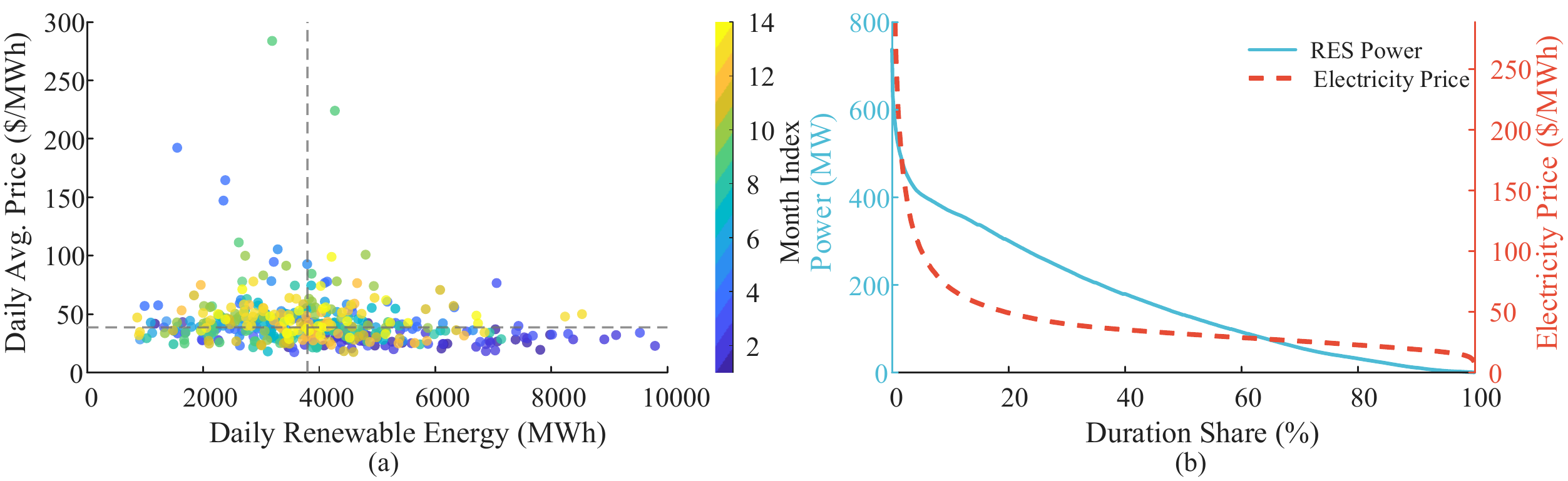}
     \caption{\rmfamily Statistical characteristics of renewable generation and electricity price data: (a) daily renewable energy versus daily average electricity price and (b) duration curves of renewable generation and real-time electricity price.}
     \captionsetup{justification=centering}\label{RES_1}
  \end{center}
  \vspace{-2.5em}
\end{figure}

\begin{figure}[pos=htbp] 
  \footnotesize\rmfamily   \setlength{\abovecaptionskip}{-0.1cm}  
    \setlength{\belowcaptionskip}{-0.1cm} 
  \begin{center}  \includegraphics[width=0.7\columnwidth]{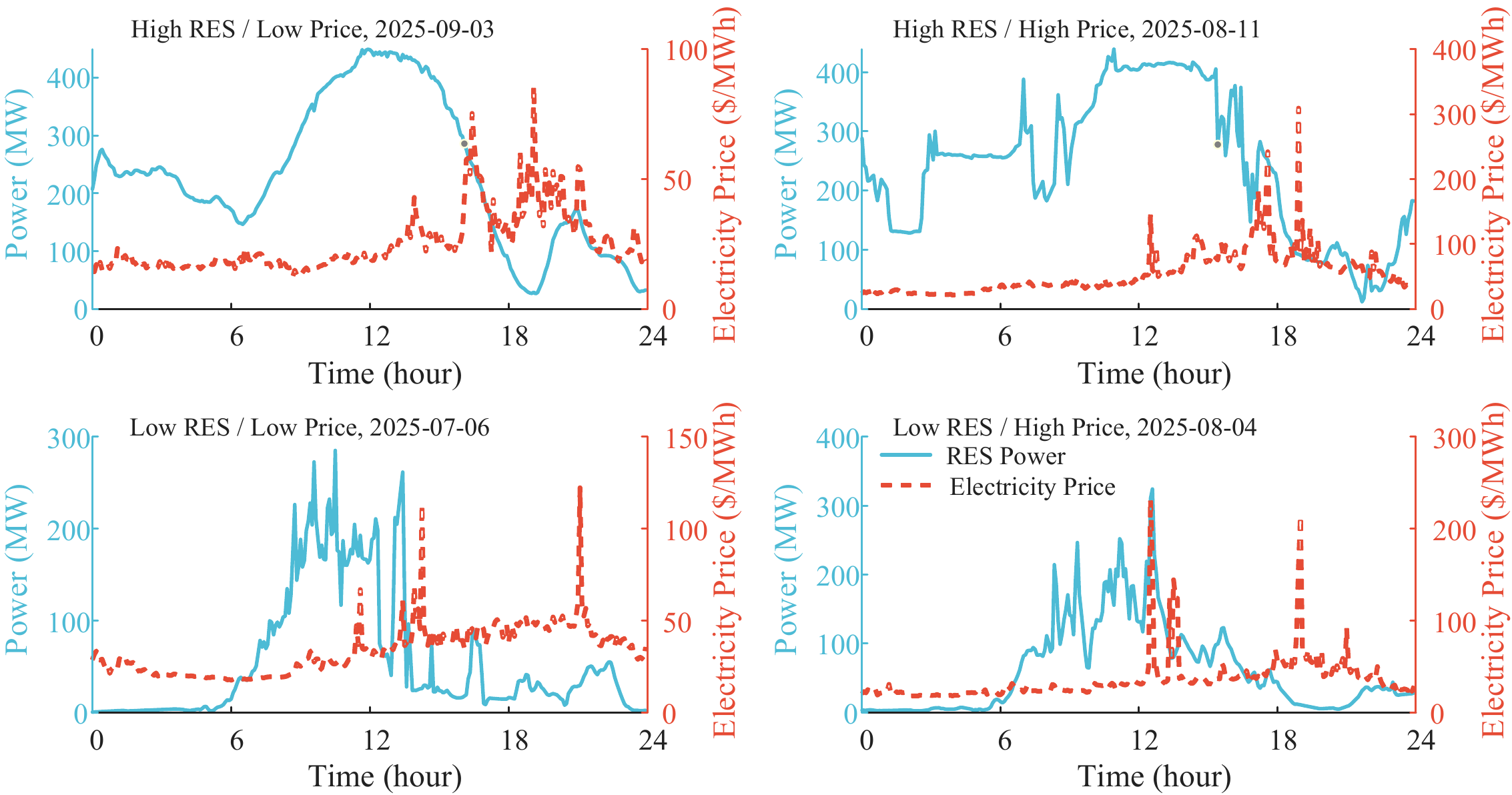}
     \caption{\rmfamily Representative daily profiles of renewable generation and real-time electricity price under four RES--price scenarios.}
     \captionsetup{justification=centering}\label{RES_2}
  \end{center}
  \vspace{-2.5em}
\end{figure}

The renewable generation data are collected from an industrial microgrid with installed wind and PV capacities of $425$ MW and $375$ MW, respectively, covering the period from Oct. 2024 to Dec. 2025. The market data are from the PJM real-time market, where the locational marginal price (LMP) is available at a $5$-min resolution. Fig.~\ref{RES_1} summarizes the statistical characteristics of renewable generation and electricity prices. Fig.~\ref{RES_1}(a) shows the relationship between daily renewable energy and daily average price, indicating that renewable availability and electricity price are not strictly synchronized. Fig.~\ref{RES_1}(b) presents the duration curves of renewable generation and real-time price, showing the uneven distribution of high-/low-renewable periods and price spikes. Fig.~\ref{RES_2} further illustrates four representative days under high-/low-RES and high-/low-price scenarios, which characterize the typical exogenous conditions considered in this study.

\subsubsection{System Configuration and Parameters}

The test system represents a short-process steel industrial microgrid with three parallel production lines, three continuous casters, and a HESS composed of BS and HS. Each heat corresponds to $80$ t of molten steel. Tables~\ref{tab:PL_parameters} and~\ref{tab:HESS_parameters} summarize the main PL and storage parameters used in the case studies.

\begin{table}[b!]
\centering
\footnotesize\rmfamily
\caption{\rmfamily Main PL and inter-stage logistics parameters of the short-process steelmaking system.}
\begin{tabular}{c p{1.1cm}<{\centering} p{1.1cm}<{\centering} p{1.1cm}<{\centering} p{1.25cm}<{\centering} p{1.25cm}<{\centering}}
\toprule
Parameter & EAF & LF & CC & EAF--LF & LF--CC \\ \midrule
Number of units & 3 & 3 & 3 & -- & -- \\
$(P^b,\underline{P},\overline{P})$ (MW) & $(60,45,75)$ & $(8,6,10)$ & $(4,4,4)$ & -- & -- \\
Energy (MWh/heat) & 34.8 & 2.4 & 2.0 & -- & -- \\
Processing time (min) & 40--60 & 20--30 & 30 & -- & -- \\
Transfer time (min) & -- & -- & -- & 5--10 & 5--10 \\
Waiting time (min) & -- & -- & -- & $\leq 10$ & $\leq 10$ \\
EAF inter-batch interval (min) & 10 & -- & -- & -- & -- \\
Daily production target (heats) & -- & -- & -- & -- & 54 \\
\bottomrule
\end{tabular}
\label{tab:PL_parameters}
\end{table}

\begin{table}[b!]
\centering
\footnotesize\rmfamily
\caption{\rmfamily Main parameters of the hybrid energy storage system.}
\begin{tabular}{c p{1.25cm}<{\centering} p{1.25cm}<{\centering} p{1.15cm}<{\centering} p{3cm}<{\centering} p{1.05cm}<{\centering} p{1.05cm}<{\centering}}
\toprule
Type & Power (MW) & Energy (MWh) & SOC range & Efficiency & Cost (\$/MWh) & Interval (min) \\ \midrule
BS & $\pm 100$ & 200 & 5--95\% & 95\%/95\% & 0.5 & 1 \\
HS & $\pm 100$ & 1000 & 5--95\% & 60--70\%/45--55\% & 0.5 & 15 \\
\bottomrule
\end{tabular}
\label{tab:HESS_parameters}
\end{table}

\subsubsection{Simulation Environment}

The simulation environment is implemented in MATLAB (R2025b) on a workstation with an Intel Core Ultra 9 275HX processor at 2.70 GHz and 32 GB RAM. Each episode represents one operating day, and the basic simulation interval is set to $1$ min. The look-ahead horizon is set to $3$ h for renewable generation and real-time electricity price signals. The dataset is divided into training and testing sets. The agents are trained by repeatedly cycling through $400$ days of historical renewable generation and price data. In both training and testing, the look-ahead RES and real-time price inputs provided to the agents are represented by $90\%$-accuracy forecast profiles, thereby emulating practical online operation with imperfect exogenous information. The realized historical profiles are used for environment evolution and reward calculation.

\subsection{Training convergence and feasibility}

\begin{figure}[pos=htbp] 
  \footnotesize\rmfamily   \setlength{\abovecaptionskip}{-0.1cm}  
    \setlength{\belowcaptionskip}{-0.1cm} 
  \begin{center}  \includegraphics[width=0.7\columnwidth]{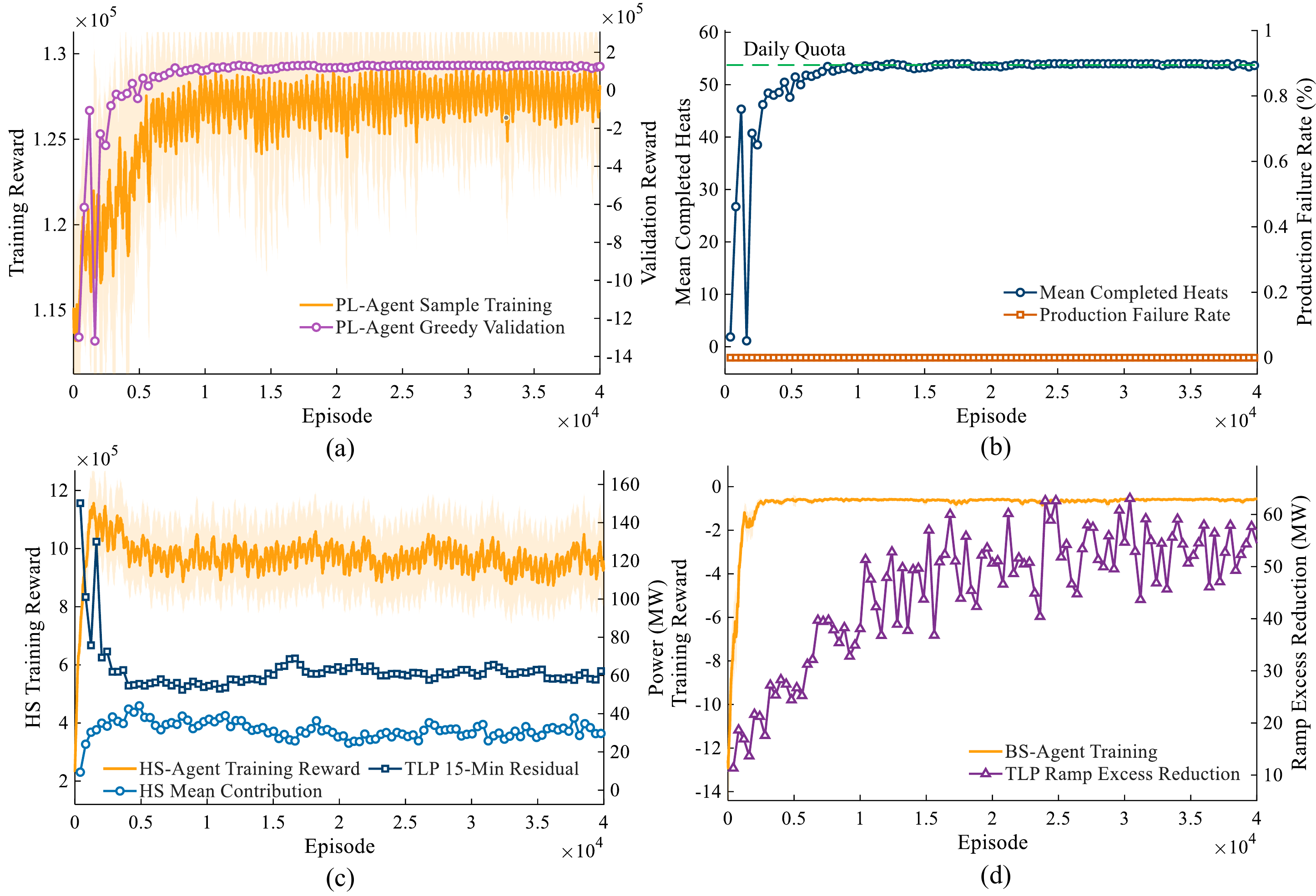}
     \caption{\rmfamily Training and greedy validation performance of the SHAC framework: (a) PL-agent sample training reward and greedy validation reward, (b) mean completed heats and production failure rate in greedy validation, (c) HS-agent training performance, (d) BS-agent training performance.}
     \captionsetup{justification=centering}\label{Training_convergence}
  \end{center}
  \vspace{-2.5em}
\end{figure}

To verify the training stability and operational feasibility of the proposed SHAC framework, Fig.~\ref{Training_convergence} presents the convergence results of different agents on training samples and validation days. As shown in Fig.~\ref{Training_convergence}(a), the training reward of the PL agent increases rapidly at the initial stage and then gradually stabilizes, remaining around $1.27\times 10^5$ in the later stage. Its greedy validation reward also converges after approximately $1.0\times 10^4$ training episodes and approaches the upper bound of the daily production reward $R_t^{\mathrm{fin}}$. This indicates that the PL agent shifts from merely advancing production to economically scheduling production in favorable periods, thereby increasing local renewable utilization and reducing grid purchase cost. Furthermore, Fig.~\ref{Training_convergence} (b) shows that the mean number of completed heats rapidly approaches the daily production target and remains around $54$ heats in the later stage, while the production failure rate remains zero throughout training. This result verifies that the safety-aware action processing and embedded process constraints effectively preserve the temporal feasibility of EAF, LF, and CC operations, avoiding invalid production transitions and process interruptions.

Fig.~\ref{Training_convergence} (c) shows the training reward of the HS agent and its regulation effect on the 15-min TLP residual. The mean HS contribution gradually increases and stabilizes at approximately $35$--$45$ MW, while the 15-min TLP residual after HS regulation decreases from more than $140$ MW at the initial stage to approximately $55$--$65$ MW. This demonstrates that the HS agent effectively mitigates the medium timescale power imbalance induced by PL scheduling through its energy buffering capability, thereby reducing the regulation burden assigned to the downstream BS agent. Fig.~\ref{Training_convergence} (d) presents the training reward of the BS agent and the TLP ramp excess reduction. As training proceeds, the BS training reward improves gradually, and the TLP ramp excess reduction increases from approximately $10$ MW at the initial stage to approximately $50$--$60$ MW before reaching a stable level.

\subsection{Operational performance of SHAC framework}

\begin{figure}[pos=htbp] 
  \footnotesize\rmfamily   \setlength{\abovecaptionskip}{-0.1cm}  
    \setlength{\belowcaptionskip}{-0.1cm} 
  \begin{center}  \includegraphics[width=0.7\columnwidth]{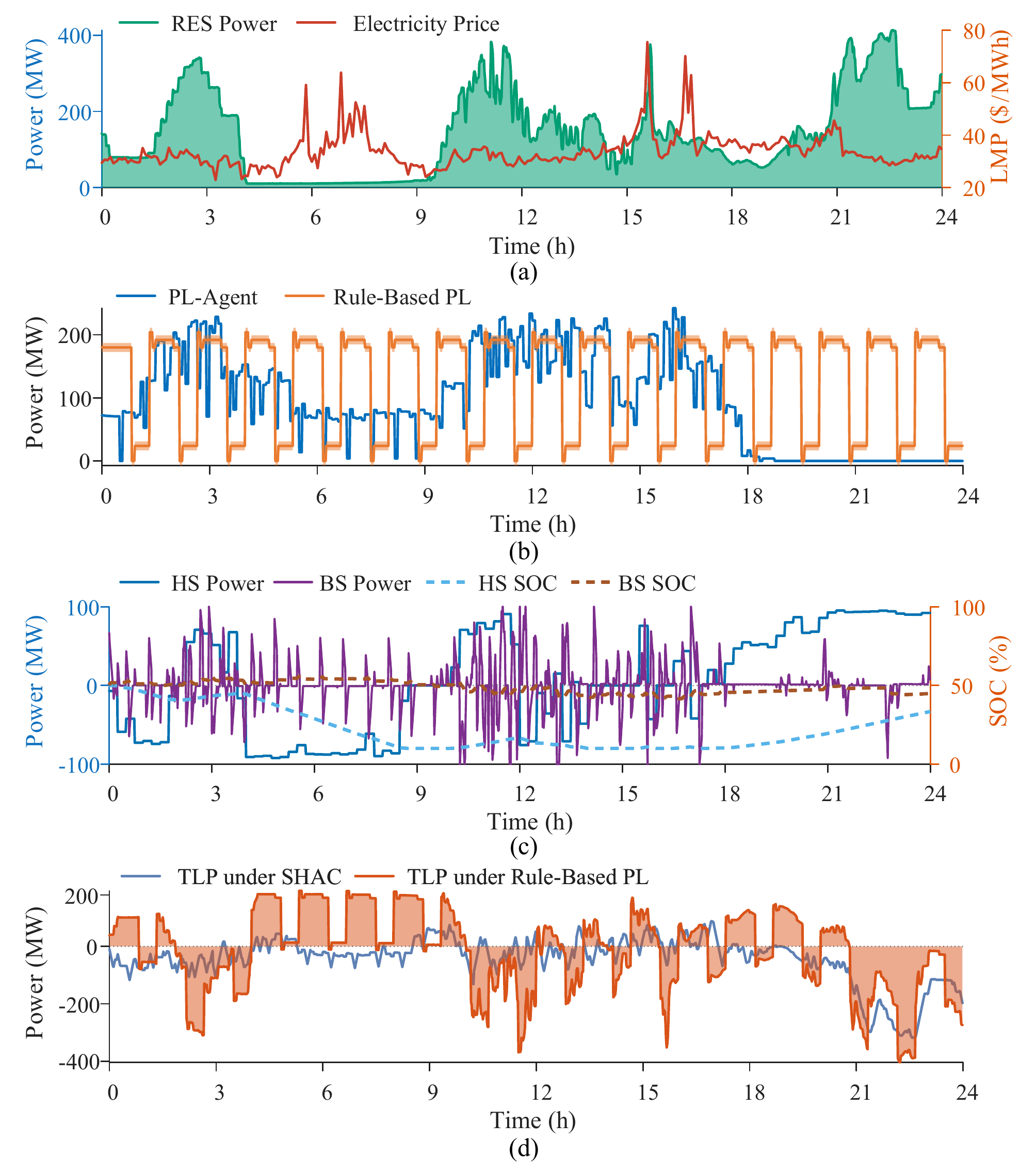}
     \caption{\rmfamily Representative day operational performance of the SHAC framework: (a) RES power and electricity price, (b) PL-agent power and rule-based PL power, (c) HS and BS power responses with SOC trajectories, (d) TLP under SHAC and rule-based PL operation.}
     \captionsetup{justification=centering}\label{Daily_Operation}
  \end{center}
  \vspace{-2.5em}
\end{figure}

To further illustrate the intraday operation mechanism of the SHAC framework, Fig. \ref{Daily_Operation} presents the multi-agent operation trajectories on a representative testing day. As shown in Fig. \ref{Daily_Operation}(a), renewable generation exhibits pronounced intraday fluctuations, while electricity prices show several peak periods. Fig. \ref{Daily_Operation}(b) compares the PL power profiles under the PL agent and the existing rule-based production strategy. The rule-based strategy starts production lines at a fixed rhythm, leading to periodic step changes in PL power and limited responsiveness to intraday variations in renewable generation and electricity prices. In contrast, the PL agent provides more flexible production scheduling and, using only a 3 h forecast horizon, jointly accounts for local renewable generation, electricity prices, and the states of the two storage systems when adjusting the production rhythm. Fig.~\ref{Daily_Operation}(c) shows the power responses and SOC trajectories of HS and BS. The HS power varies relatively smoothly and mainly provides medium timescale energy buffering, whereas the BS power responds more frequently to compensate minute-level fast fluctuations. The SOCs of both storage systems remain within feasible ranges, indicating that the terminal SOC reward helps maintain storage feasibility while supporting TLP regulation. Fig.~\ref{Daily_Operation}(d) compares the TLP trajectories under the full SHAC strategy and the rule-based production strategy. Under rule-based production, the fixed step changes in PL power lead to pronounced peak-valley variations and rapid ramping in TLP. In contrast, SHAC effectively mitigates the TLP impacts caused by the superposition of production load variations and renewable fluctuations through the sequential coordination of PL, HS, and BS, resulting in a smoother TLP trajectory.

\begin{figure}[pos=htbp] 
  \footnotesize\rmfamily   \setlength{\abovecaptionskip}{-0.1cm}  
    \setlength{\belowcaptionskip}{-0.1cm} 
  \begin{center}  \includegraphics[width=0.9\columnwidth]{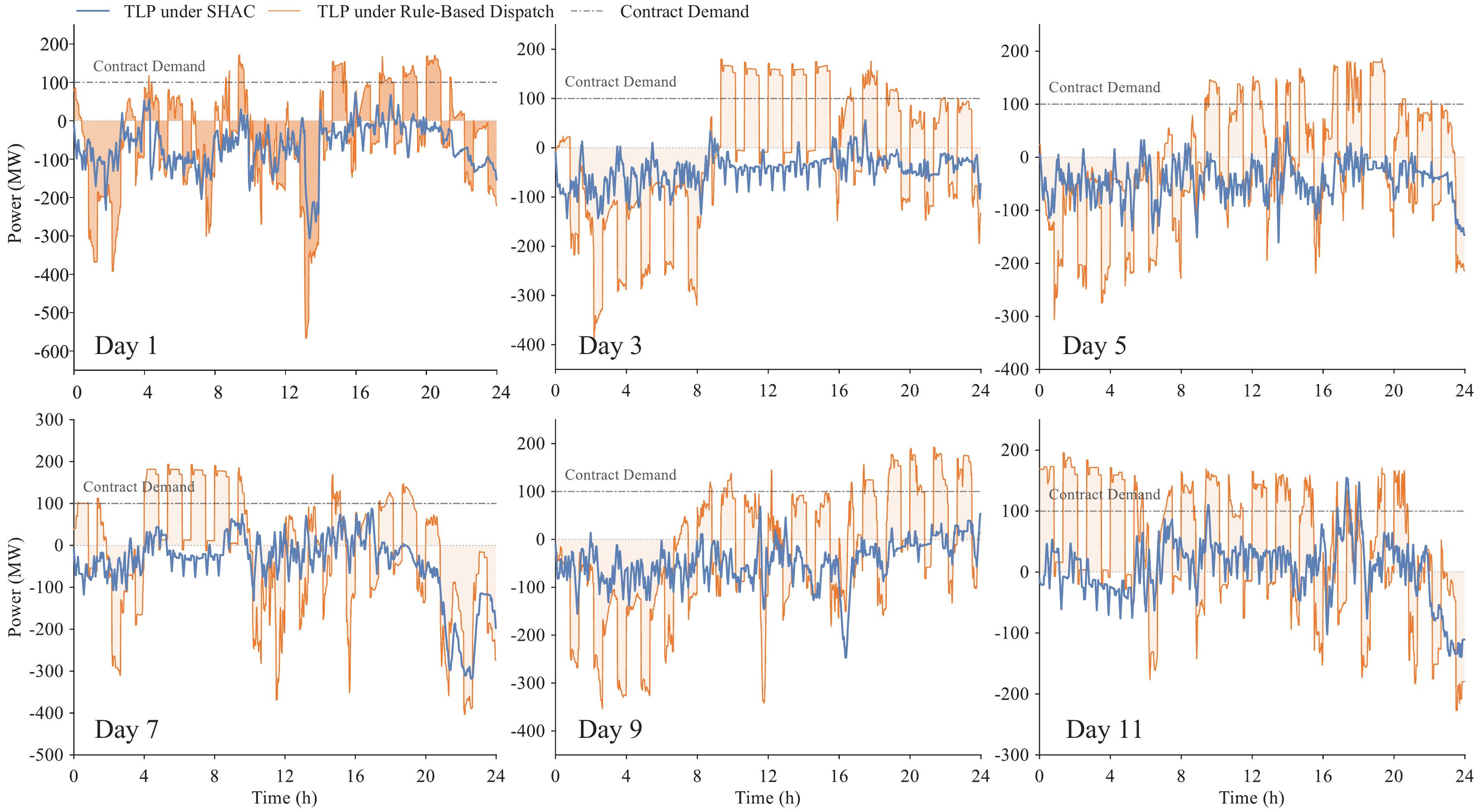}
     \caption{\rmfamily TLP shaping performance of the SHAC framework on various RES--price scenarios.}
     \captionsetup{justification=centering}\label{Days_results}
  \end{center}
  \vspace{-2.5em}
\end{figure}

Fig.~\ref{Days_results} illustrates the TLP regulation performance of SHAC on different testing days. Compared with the rule-based production strategy, SHAC substantially attenuates the peak-valley variations of the grid-interaction power in most scenarios and reduces the periods in which the contract demand is exceeded. Table~\ref{tlp_performance} reports the average results over the full testing set. SHAC reduces the total grid purchase cost, peak grid import, and cumulative time above the contract demand by $91.27\%$, $61.39\%$, and $98.64\%$, respectively. In addition, the 15-min residual, mean 1-min ramp, ramp violation count, and ramp excess are reduced by $42.53\%$, $37.45\%$, $89.25\%$, and $96.91\%$, respectively. These results demonstrate that SHAC improves grid-interaction performance in terms of operating economy, contract demand compliance, TLP smoothness, and reduced grid export.

\begin{table*}[pos=htbp] 
\vspace{-2.2em}
\centering
\footnotesize\rmfamily
\setlength{\abovecaptionskip}{0pt}
\setlength{\belowcaptionskip}{0em}
\caption{\rmfamily Average grid-friendly TLP performance over all testing days.}
\begin{threeparttable}
\setlength{\tabcolsep}{3.0pt}
\renewcommand{\arraystretch}{1.12}
\begin{tabular}{c p{0.95cm}<{\centering} p{0.95cm}<{\centering} p{0.95cm}<{\centering} p{0.95cm}<{\centering} p{0.95cm}<{\centering} p{1.0cm}<{\centering} p{0.95cm}<{\centering} p{0.95cm}<{\centering} p{1.0cm}<{\centering}}
\toprule
Model
& $C^{\mathrm{base}}$
& $C^{\mathrm{dem}}$
& $C^{\mathrm{grd}}$
& $P_{\max}^{\mathrm{imp}}$
& $T^{\mathrm{dem}}$
& $\bar P_{\mathrm{res}}^{15}$
& $\bar R^{1}$
& $N^{\mathrm{ramp}}$
& $E^{\mathrm{ramp}}$
\\
& ($\$10^4$)
& ($\$10^4$)
& ($\$10^4$)
& (MW)
& (h)
& (MW)
& (MW)
&
& (MW)
\\
\midrule
Original TLP
& 3.02
& 1.61
& 4.63
& 181.79
& 4.56
& 107.26
& 9.69
& 283.47
& 9115.50
\\
Grid-friendly TLP under SHAC
& 0.40
& 0.01
& 0.40
& 70.19
& 0.06
& 61.65
& 6.06
& 30.47
& 281.26
\\
Reduction rate
& 86.91\%
& 99.48\%
& 91.27\%
& 61.39\%
& 98.64\%
& 42.53\%
& 37.45\%
& 89.25\%
& 96.91\%
\\
\bottomrule
\end{tabular}
\begin{tablenotes}
\footnotesize
\item $C^{\mathrm{base}}$: regular grid purchase cost within $D^{\mathrm{dem}}$; 
$C^{\mathrm{dem}}$: contract demand exceedance cost; 
$C^{\mathrm{grd}}$: total grid purchase cost; 
$P_{\max}^{\mathrm{imp}}$: peak grid import; 
$T^{\mathrm{dem}}$: cumulative time above $D^{\mathrm{dem}}$; 
$\bar P_{\mathrm{res}}^{15}$: mean absolute 15-min TLP residual; 
$\bar R^{1}$: mean 1-min TLP ramp; 
$N^{\mathrm{ramp}}$: ramp violation count; 
$E^{\mathrm{ramp}}$: cumulative ramp excess.
\end{tablenotes}
\end{threeparttable}
\label{tlp_performance}
\vspace{-2.2em}
\end{table*}

\subsection{Ablation Studies}

To further examine the effectiveness of the key structural designs in SHAC, two ablation variants are constructed for comparison. Shared-Critic SHAC (SC-SHAC) removes the role-specific credit assignment mechanism by replacing the role-specific critics and rewards of PL, HS, and BS with a shared critic and a unified reward. Parallel SHAC (P-SHAC) preserves the role-specific critics and rewards, but removes the sequential PL $\rightarrow$ HS $\rightarrow$ BS conditioning structure, such that the three agents make decisions in parallel based on their own observations.

\begin{figure}[pos=htbp] 
  \footnotesize\rmfamily   \setlength{\abovecaptionskip}{-0.1cm}  
    \setlength{\belowcaptionskip}{-0.1cm} 
  \begin{center}  \includegraphics[width=0.5\columnwidth]{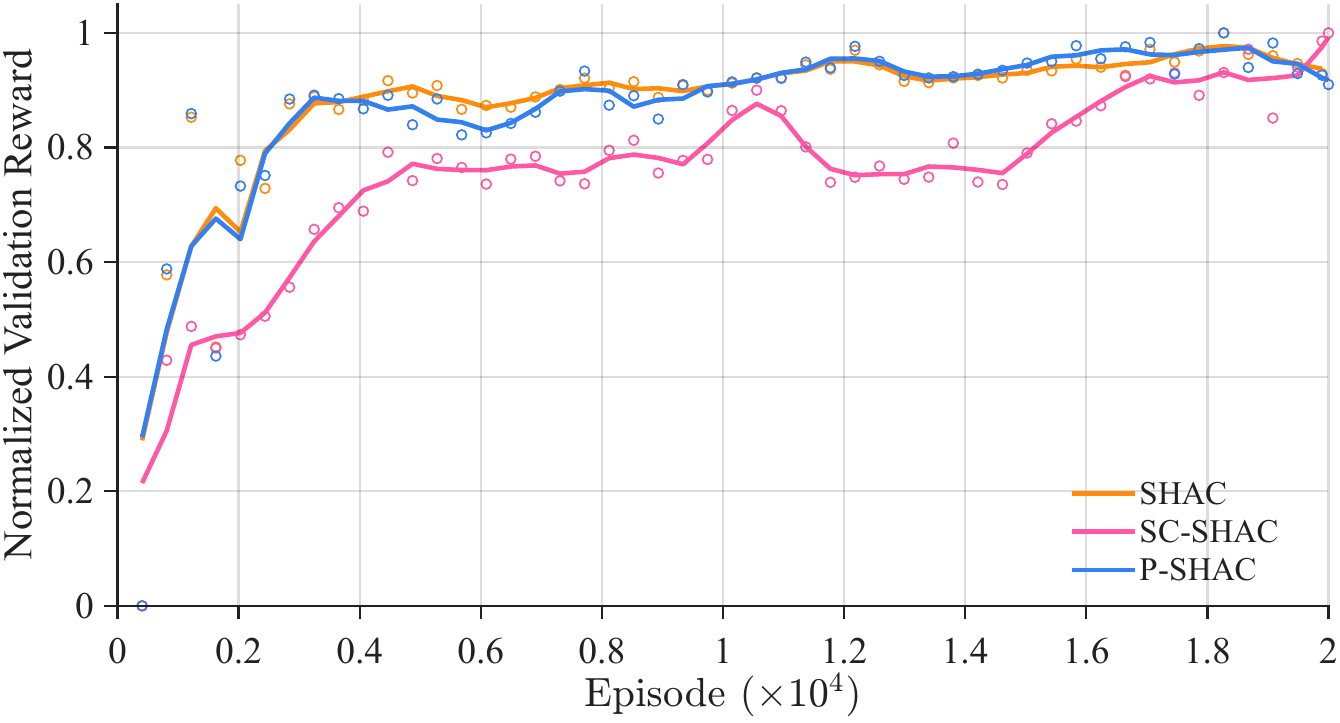}
     \caption{\rmfamily Normalized diagnostic validation reward trends of SHAC, SC-SHAC, and P-SHAC during training.}
     \captionsetup{justification=centering}\label{Ablation_validation_whole}
  \end{center}
  \vspace{-2.5em}
\end{figure}

\begin{figure}[pos=htbp] 
  \footnotesize\rmfamily   \setlength{\abovecaptionskip}{-0.1cm}  
    \setlength{\belowcaptionskip}{-0.1cm} 
  \begin{center}  \includegraphics[width=0.5\columnwidth]{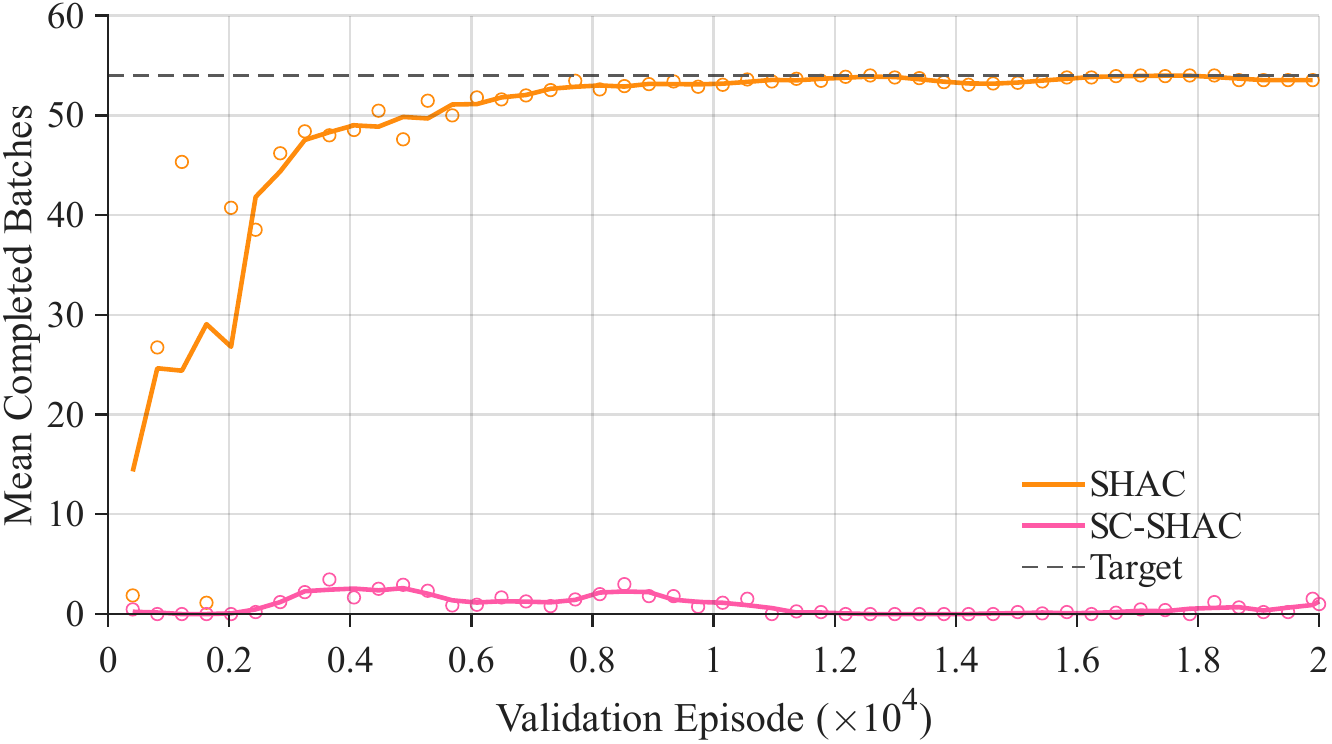}
     \caption{\rmfamily Mean completed production heats during validation for SHAC and SC-SHAC.}
     \captionsetup{justification=centering}\label{Ablation_PL}
  \end{center}
  \vspace{-2.5em}
\end{figure}

\begin{figure}[pos=htbp] 
  \footnotesize\rmfamily   \setlength{\abovecaptionskip}{-0.1cm}  
    \setlength{\belowcaptionskip}{-0.1cm} 
  \begin{center}  \includegraphics[width=0.5\columnwidth]{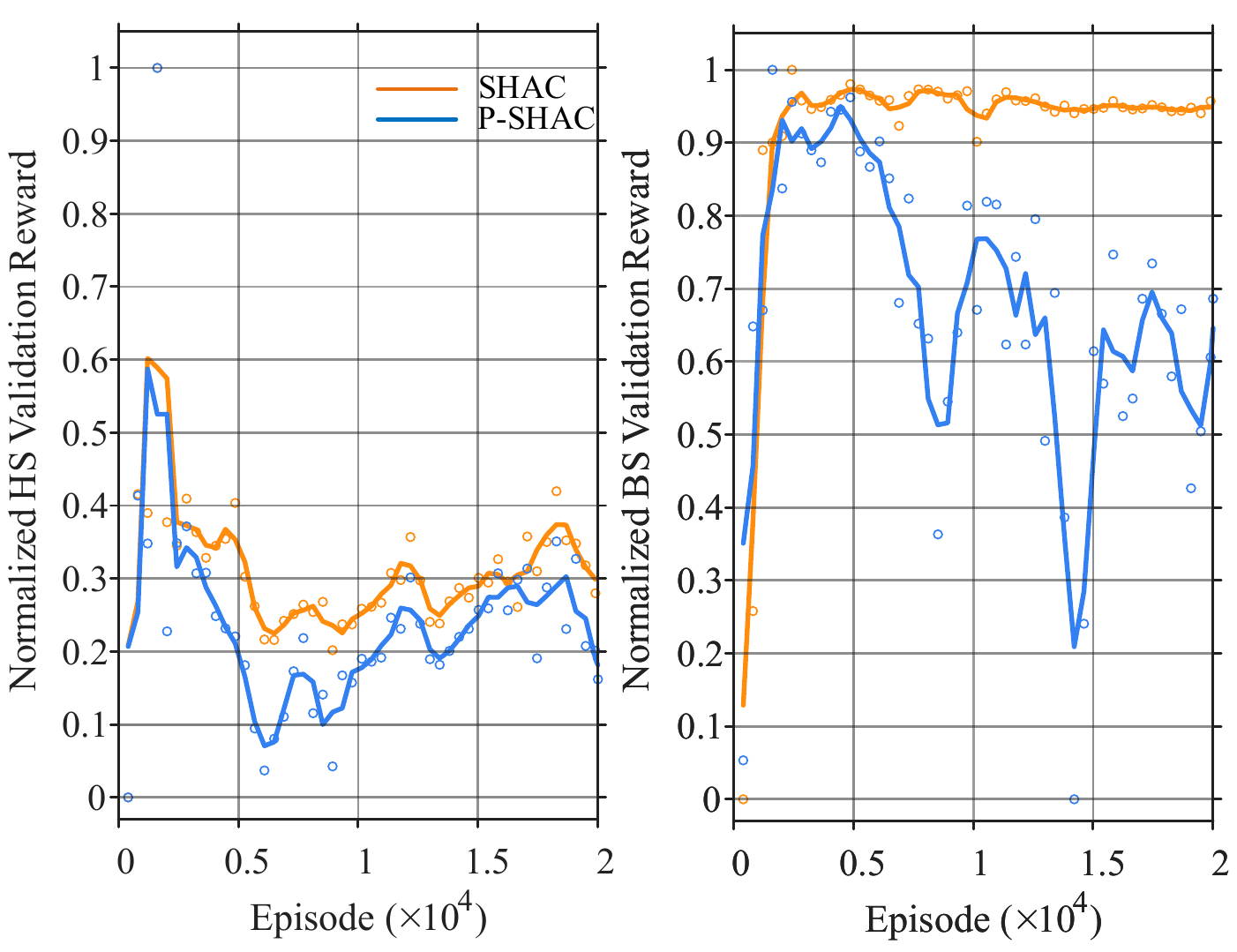}
     \caption{\rmfamily HS- and BS-agent validation reward comparison between SHAC and P-SHAC.}
     \captionsetup{justification=centering}\label{Ablation_validation_seperate}
  \end{center}
  \vspace{-2.5em}
\end{figure}

Fig.~\ref{Ablation_validation_whole} presents the normalized diagnostic validation rewards of the three methods during training. SHAC, SC-SHAC, and P-SHAC all exhibit clear reward improvement trends, indicating that the sample-based training processes are effective for all three schemes and that each variant develops a decision logic consistent with its own structural design. Fig.~\ref{Ablation_PL} further reveals the key limitation of SC-SHAC. Although its normalized validation reward increases in the later training stage, the mean number of completed daily production heats remains far below the daily quota target. This result indicates that the shared critic and unified reward cannot provide a sufficiently clear long-term production credit signal for the PL agent, and may induce policies that apparently improve grid-side rewards while failing to satisfy the production requirement. Fig.~\ref{Ablation_validation_seperate} compares the role-specific validation rewards of SHAC and P-SHAC. The overall learning trends of the two methods are relatively close, but P-SHAC exhibits stronger fluctuations in the HS- and BS-agent rewards, with a particularly evident degradation in the stability of the BS reward in the later training stage. This suggests that the main function of sequential conditioning is not to alter the production learning capability of the PL agent, but to enhance the ability of downstream storage agents to perceive and compensate for the residual after PL and HS actions. Therefore, role-specific credit assignment determines whether SHAC can learn an effective scheduling policy that satisfies production constraints, while the sequential update further improves the coordination stability of HS and BS in residual regulation and ramp mitigation.

\subsection{Comparative performance evaluation}

To further evaluate the advantages of SHAC in grid-friendly TLP shaping, it is compared with three representative coordination methods: rule-based control (RB-Control), hierarchical reference tracking control (HRTC), and multi-timescale model predictive control (MT-MPC). RB-Control completes the daily production task following a fixed and uniform production rhythm, and coordinates HS and BS through rule-based power compensation. HRTC represents a conventional hierarchical reference tracking control method, where a smooth TLP reference trajectory is first generated, and PL, HS, and BS are then coordinated to track this reference over a 3 h prediction horizon. MT-MPC adopts a hierarchical multi-objective rolling optimization framework, which jointly considers grid purchase cost, contract demand exceedance, 15-min TLP residual, ramp excess, and storage operating cost within a 3 h prediction horizon, and is solved by YALMIP/Gurobi. All comparative methods use the same renewable generation and electricity price data as SHAC, and online decisions are made under a 90\% forecast accuracy setting. It should be noted that constructing a full-day global MILP requires simultaneously modeling 54 heats, all process units, and storage dynamics. As the number of time steps, devices, and batches increases, the number of binary variables grows rapidly, making the global MILP difficult to solve. Therefore, MT-MPC is adopted as a computationally tractable optimization-based benchmark.

\begin{figure}[pos=htbp] 
  \footnotesize\rmfamily   \setlength{\abovecaptionskip}{-0.1cm}  
    \setlength{\belowcaptionskip}{-0.1cm} 
  \begin{center}  \includegraphics[width=0.6\columnwidth]{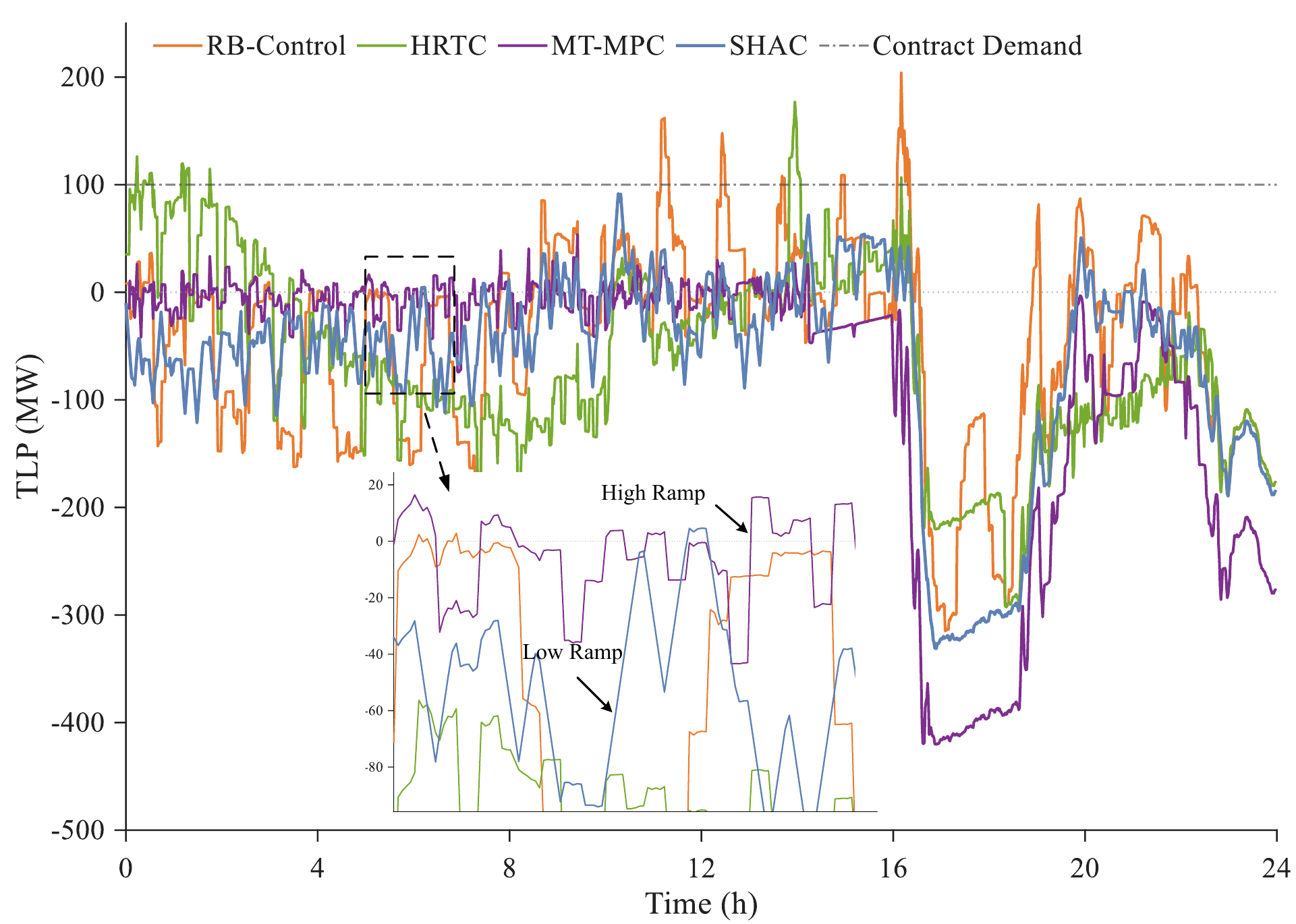}
     \caption{\rmfamily Comparative TLP shaping performance of different coordination methods on a representative testing day.}
     \captionsetup{justification=centering}\label{Compare}
  \end{center}
  \vspace{-2.5em}
\end{figure}

Fig.~\ref{Compare} presents the TLP trajectories of different methods on a representative testing day. Due to its fixed production rhythm, RB-Control leads to pronounced peak-valley variations and rapid ramping in TLP. HRTC improves the TLP profile in some periods through reference tracking, however its performance depends on the predefined reference and is less adaptive to the dynamic coupling between renewable generation and process loads. MT-MPC can substantially reduce grid purchase in some periods, however, under the limited 3-h prediction horizon and forecast errors, it cannot simultaneously achieve the best performance in grid purchase cost, 15-min residual, ramp mitigation, and reverse power control. In contrast, SHAC achieves more stable dynamic regulation through the sequential heterogeneous coordination of PL, HS, and BS, without relying on a predefined TLP reference.

Table~\ref{comparative_tlp_performance} further reports the statistical results over all testing days. Although the total grid purchase cost of SHAC is higher than that of MT-MPC, SHAC achieves the lowest 15-min residual, ramp violation count, and reverse energy. Its online decision time is only 0.4 ms per 1-min step, which is significantly lower than those of HRTC and MT-MPC. These results indicate that SHAC achieves a more balanced grid-friendly TLP shaping performance with a low computational burden, making it particularly suitable for real-time dispatch of industrial microgrids with multiple discrete process loads and multi-timescale storage resources.

\begin{table*}[pos=htbp] 
\centering
\footnotesize\rmfamily
\caption{\rmfamily Grid-friendly TLP performance of different coordination methods over all testing days.}
\begin{threeparttable}
\setlength{\tabcolsep}{6.0pt}
\renewcommand{\arraystretch}{1.12}
\begin{tabular}{c p{1.5cm}<{\centering} p{1.5cm}<{\centering} p{1.5cm}<{\centering} p{1.5cm}<{\centering} p{2cm}<{\centering}}
\toprule
Model
& $C^{\mathrm{grd}}$
& $\bar P_{\mathrm{res}}^{15}$
& $N^{\mathrm{ramp}}$
& $E^{\mathrm{rev}}$
& Decision Time
\\
& ($\$10^4$)
& (MW)
&
& (MWh)
& /1-min step
\\
\midrule
RB-Control
& 0.99
& 67.65
& 249.63
& 1982.2  
& 0.1 ms
\\
HRTC
& 0.81
& 85.45
& 202.65
& 1107.6
& 2.2 s
\\
MT-MPC
& 0.34
& 77.81
& 167.42
& 1486.9
& 18.7 s
\\
SHAC
& 0.40
& 61.65
& 30.47
& 1055.1
& 0.4 ms
\\
\bottomrule
\end{tabular}
\end{threeparttable}
\label{comparative_tlp_performance}
\end{table*}

\section{Conclusion}~\label{conclusion}

This paper investigated the grid-friendly TLP shaping problem for short-process steelmaking industrial microgrids and proposed a reference-free sequential heterogeneous-agent coordination framework, namely SHAC. In this framework, process loads, hydrogen storage, and battery storage are modeled as heterogeneous agents with different physical responsibilities, action structures, and response timescales. Through sequential coordination, role-specific credit assignment, and safety-aware action processing, the proposed framework enables online coordinated dispatch of multi-type flexible resources under strict process constraints.

Numerical results based on real renewable generation and electricity market data demonstrate that SHAC can achieve stable 1-min online decision-making without relying on a predefined TLP reference trajectory, while satisfying the production feasibility requirements of short-process steelmaking. Compared with the original operation, the proposed method significantly reduces the total grid purchase cost, contract-demand exceedance time, and cumulative ramp excess, demonstrating its comprehensive improvements in operational economy, contract-demand management, and TLP smoothing. The ablation studies further verify the necessity of role-specific credit assignment for production-task completion and policy stability, as well as the contribution of sequentially conditioned decision-making to the coordinated residual regulation of HS and BS. Compared with rule-based control, hierarchical reference tracking control, and multi-timescale MPC, SHAC achieves a more balanced grid-friendly TLP shaping performance while maintaining millisecond-level online computational efficiency.

Future work will further consider internal network constraints, communication delays, and coordinated operation of multiple industrial microgrids to improve the applicability of the proposed framework in more complex engineering scenarios.

\printcredits
\vspace{0.3cm}

\noindent \textbf{Declaration of competing interest}

The authors declare that they have no known competing financial interests or personal relationships that could have appeared to influence the work reported in this paper.

\vspace{0.3cm}

\noindent \textbf{Data availability}

The electricity price data can be downloaded from~\cite{data}.

\vspace{0.2cm}

\noindent \textbf{Acknowledgements}

This work was supported by Science and Technology Program of Xinjiang Uyghur Autonomous Region (2024B01001-2).

\vspace{-0.2cm}

\bibliographystyle{unsrt}

\bibliography{cas-refs-1}

@misc{data,
  author       = {{PJM Interconnection, L.L.C.}},
  title        = {{Data Miner 2: Real-Time Five Minute LMPs}},
  year         = {2026},
  howpublished = {\url{https://dataminer2.pjm.com/feed/rt_fivemin_hrl_lmps/definition}},
}

@article{xi2026industrialgreenmicrogrids,
  title={Toward sustainable and low-carbon industrial transformation: Insights from industrial green microgrids},
  author={Xi, Yufei and Wang, Xinyi and Cheng, Lin and Chen, Meng and Lestas, Ioannis and Shi, Fashun and Huang, Kaidi},
  journal={Applied Energy},
  volume={417},
  pages={128055},
  year={2026},
  doi={10.1016/j.apenergy.2026.128055},
  publisher={Elsevier}
}

@article{zhang2025industrialflexibility,
  title={A Review of Industrial Load Flexibility Enhancement for Demand-Response Interaction},
  author={Zhang, Jiubo and Zhou, Bowen and Yang, Zhile and Guo, Yuanjun and Lv, Chen and Xu, Xiaofeng and Liu, Jichun},
  journal={Sustainability},
  volume={17},
  number={11},
  pages={4938},
  year={2025},
  doi={10.3390/su17114938},
  publisher={MDPI}
}

@article{giovanniello2023hybrid,
  title={Hybrid lithium-ion battery and hydrogen energy storage systems for a wind-supplied microgrid},
  author={Giovanniello, Michael Anthony and Wu, Xiao-Yu},
  journal={Applied Energy},
  volume={345},
  pages={121311},
  year={2023},
  doi={10.1016/j.apenergy.2023.121311},
  publisher={Elsevier}
}

@article{su2023,
  title={Multi-objective scheduling of a steelmaking plant integrated with renewable energy sources and energy storage systems: Balancing costs, emissions and make-span},
  author={Su, Pengfei and Zhou, Yue and Wu, Jianzhong},
  journal={Journal of Cleaner Production},
  volume={428},
  pages={139350},
  year={2023},
  publisher={Elsevier}
}

@article{qi2025,
  title={Long-term energy management for microgrid with hybrid hydrogen-battery energy storage: A prediction-free coordinated optimization framework},
  author={Qi, Ning and Huang, Kaidi and Fan, Zhiyuan and Xu, Bolun},
  journal={Applied Energy},
  volume={377},
  pages={124485},
  year={2025},
  publisher={Elsevier}
}

@article{PL1,
  title={A demand response energy management scheme for industrial facilities in smart grid},
  author={Ding, Yue Min and Hong, Seung Ho and Li, Xiao Hui},
  journal={IEEE transactions on industrial informatics},
  volume={10},
  number={4},
  pages={2257--2269},
  year={2014},
  publisher={IEEE}
}

@article{PL2,
  title={Optimal industrial load control in smart grid},
  author={Gholian, Armen and Mohsenian-Rad, Hamed and Hua, Yingbo},
  journal={IEEE Transactions on Smart Grid},
  volume={7},
  number={5},
  pages={2305--2316},
  year={2015},
  publisher={IEEE}
}

@article{PL3,
  title={A real-time decision model for industrial load management in a smart grid},
  author={Yu, Mengmeng and Lu, Renzhi and Hong, Seung Ho},
  journal={Applied energy},
  volume={183},
  pages={1488--1497},
  year={2016},
  publisher={Elsevier}
}

@article{PL4,
  title={Cost-effective scheduling of steel plants with flexible EAFs},
  author={Zhang, Xiao and Hug, Gabriela and Harjunkoski, Iiro},
  journal={IEEE Transactions on Smart Grid},
  volume={8},
  number={1},
  pages={239--249},
  year={2016},
  publisher={IEEE}
}

@article{PL5,
  title={Demand response of ancillary service from industrial loads coordinated with energy storage},
  author={Zhang, Xiao and Hug, Gabriela and Kolter, J Zico and Harjunkoski, Iiro},
  journal={IEEE Transactions on Power Systems},
  volume={33},
  number={1},
  pages={951--961},
  year={2017},
  publisher={IEEE}
}

@article{PL6,
  title={Demand-side management via optimal production scheduling in power-intensive industries: The case of metal casting process},
  author={Ramin, Danial and Spinelli, Stefano and Brusaferri, Alessandro},
  journal={Applied Energy},
  volume={225},
  pages={622--636},
  year={2018},
  publisher={Elsevier}
}

@article{PL7,
  title={Robust self-scheduling of operational processes for industrial demand response aggregators},
  author={Golmohamadi, Hessam and Keypour, Reza and Bak-Jensen, Birgitte and Pillai, Jayakrishnan R and Khooban, Mohammad Hassan},
  journal={IEEE Transactions on Industrial Electronics},
  volume={67},
  number={2},
  pages={1387--1395},
  year={2019},
  publisher={IEEE}
}

@article{PL8,
  title={Quantifying flexibility provisions of the ladle furnace refining process as cuttable loads in the iron and steel industry},
  author={Wang, Jiayang and Wang, Qiang and Sun, Wenqiang},
  journal={Applied Energy},
  volume={342},
  pages={121178},
  year={2023},
  publisher={Elsevier}
}

@article{PL9,
  title={Cost-effective scheduling of a hydrogen-based iron and steel plant powered by a grid-assisted renewable energy system},
  author={Su, Pengfei and Zhou, Yue and Li, Hongyi and Perez, Hector D and Wu, Jianzhong},
  journal={Applied Energy},
  volume={384},
  pages={125412},
  year={2025},
  publisher={Elsevier}
}

@article{PL10,
  title={Distributionally robust chance-constrained energy management of steel industrial microgrid with energy storage in distribution market},
  author={Fu, Linbo and Chen, Houhe and Zhang, Rufeng and Jiang, Tao and Li, Guoqing and Qu, Rui},
  journal={Applied Energy},
  volume={400},
  pages={126574},
  year={2025},
  publisher={Elsevier}
}

@article{PL11,
  title={Efficient scheduling of discrete industrial processes through continuous modeling},
  author={Lyu, Ruike and Su, Xiangbo and Du, Ershun and Guo, Hongye and Chen, Qixin and Kang, Chongqing},
  journal={IEEE Transactions on Smart Grid},
  year={2025},
  publisher={IEEE}
}

@article{TLP1,
  title={Smoothing tie-line power fluctuations for industrial microgrids by demand side control: An output regulation approach},
  author={Zhang, Chen and Lin, Wei and Ke, Deping and Sun, Yuanzhang},
  journal={IEEE Transactions on Power Systems},
  volume={34},
  number={5},
  pages={3716--3728},
  year={2019},
  publisher={IEEE}
}

@article{TLP2,
  title={Suppressing active power fluctuations at PCC in grid-connection microgrids via multiple BESSs: A collaborative multi-agent reinforcement learning approach},
  author={He, Wangli and Li, Chengyuan and Cai, Chenhao and Qing, Xiangyun and Du, Wenli},
  journal={Applied Energy},
  volume={373},
  pages={123858},
  year={2024},
  publisher={Elsevier}
}

@article{TLP3,
  title={Multi-time-scale energy management of renewable microgrids considering grid-friendly interaction},
  author={Li, Shenglin and Zhu, Jizhong and Dong, Hanjiang and Zhu, Haohao and Luo, Fengji and Borghetti, Alberto},
  journal={Applied Energy},
  volume={367},
  pages={123428},
  year={2024},
  publisher={Elsevier}
}

@article{TLP4,
  title={Flat tie-line power scheduling control of grid-connected hybrid microgrids},
  author={Xiao, Zhao-xia and Guerrero, Josep M and Shuang, Jia and Sera, Dezso and Schaltz, Erik and V{\'a}squez, Juan C},
  journal={Applied Energy},
  volume={210},
  pages={786--799},
  year={2018},
  publisher={Elsevier}
}

@article{TLP5,
  title={A demand response and battery storage coordination algorithm for providing microgrid tie-line smoothing services},
  author={Wang, Dan and Ge, Shaoyun and Jia, Hongjie and Wang, Chengshan and Zhou, Yue and Lu, Ning and Kong, Xiangyu},
  journal={IEEE Transactions on Sustainable Energy},
  volume={5},
  number={2},
  pages={476--486},
  year={2014},
  publisher={IEEE}
}

@article{TLP6,
  title={Data center holistic demand response algorithm to smooth microgrid tie-line power fluctuation},
  author={Yang, Ting and Zhao, Yingjie and Pen, Haibo and Wang, Zhaoxia},
  journal={Applied Energy},
  volume={231},
  pages={277--287},
  year={2018},
  publisher={Elsevier}
}

@article{TLP7,
  title={Multi-data center tie-line power smoothing method based on demand response},
  author={Yang, Ting and Hou, Yuxing and Cai, Shaotang and Yu, Jie and Pen, Haibo},
  journal={IEEE Transactions on Cloud Computing},
  volume={12},
  number={4},
  pages={983--995},
  year={2024},
  publisher={IEEE}
}

@article{TLP8,
  title={A novel rolling optimization strategy considering grid-connected power fluctuations smoothing for renewable energy microgrids},
  author={Li, Shenglin and Zhu, Jizhong and Dong, Hanjiang and Zhu, Haohao and Fan, Junwei},
  journal={Applied Energy},
  volume={309},
  pages={118441},
  year={2022},
  publisher={Elsevier}
}

@article{TLP9,
  title={A coordinated multitimescale model predictive control for output power smoothing in hybrid microgrid incorporating hydrogen energy storage},
  author={Abdelghany, Muhammad Bakr and Al-Durra, Ahmed and Zeineldin, Hatem H and Gao, Fei},
  journal={IEEE Transactions on Industrial Informatics},
  volume={20},
  number={9},
  pages={10987--11001},
  year={2024},
  publisher={IEEE}
}

@misc{zhinan,
  author       = {{Ministry of Industry and Information Technology of the People's Republic of China et al.}},
  title        = {{Guidelines for the Construction and Application of Industrial Green Microgrids (2026--2030)}},
  year         = {2026},
  howpublished = {\url{https://www.miit.gov.cn/zwgk/zcwj/wjfb/tz/art/2026/art_44c5364d80b748e4b797bc115388c6aa.html}},
}

@article{MARL1,
  title={Multi-agent deep reinforcement learning based demand response for discrete manufacturing systems energy management},
  author={Lu, Renzhi and Li, Yi-Chang and Li, Yuting and Jiang, Junhui and Ding, Yuemin},
  journal={Applied Energy},
  volume={276},
  pages={115473},
  year={2020},
  publisher={Elsevier}
}

@article{MARL2,
  title={Multi-agent deep reinforcement learning based demand response and energy management for heavy industries with discrete manufacturing systems},
  author={Bashyal, Atit and Boroukhian, Tina and Veerachanchai, Pakin and Naransukh, Myanganbayar and Wicaksono, Hendro},
  journal={Applied Energy},
  volume={392},
  pages={125990},
  year={2025},
  publisher={Elsevier}
}

@article{MARL3,
  title={Energy management based on multi-agent deep reinforcement learning for a multi-energy industrial park},
  author={Zhu, Dafeng and Yang, Bo and Liu, Yuxiang and Wang, Zhaojian and Ma, Kai and Guan, Xinping},
  journal={Applied Energy},
  volume={311},
  pages={118636},
  year={2022},
  publisher={Elsevier}
}

@article{MARL4,
  title={Coordination for multienergy microgrids using multiagent reinforcement learning},
  author={Qiu, Dawei and Chen, Tianyi and Strbac, Goran and Bu, Shengrong},
  journal={IEEE Transactions on Industrial Informatics},
  volume={19},
  number={4},
  pages={5689--5700},
  year={2022},
  publisher={IEEE}
}

@article{MARL5,
  title={Physics-model-free heat-electricity energy management of multiple microgrids based on surrogate model-enabled multi-agent deep reinforcement learning},
  author={Li, Sichen and Hu, Weihao and Cao, Di and Chen, Zhe and Huang, Qi and Blaabjerg, Frede and Liao, Kaiji},
  journal={Applied Energy},
  volume={346},
  pages={121359},
  year={2023},
  publisher={Elsevier}
}

@article{MARL6,
  title={Renewable energy integration and microgrid energy trading using multi-agent deep reinforcement learning},
  author={Harrold, Daniel JB and Cao, Jun and Fan, Zhong},
  journal={Applied Energy},
  volume={318},
  pages={119151},
  year={2022},
  publisher={Elsevier}
}

@article{MARL7,
  title={Multi-agent reinforcement learning for energy management in microgrids with shared hydrogen storage},
  author={Toquica, David and Agbossou, Kodjo and Henao, Nilson},
  journal={International Journal of Hydrogen Energy},
  volume={144},
  pages={1019--1027},
  year={2025},
  publisher={Elsevier}
}

@article{MARL8,
  title={Multi-agent hierarchical reinforcement learning for energy management},
  author={Jendoubi, Imen and Bouffard, Fran{\c{c}}ois},
  journal={Applied Energy},
  volume={332},
  pages={120500},
  year={2023},
  publisher={Elsevier}
}

@article{MARL9,
  title={Collaborative optimization of multi-energy multi-microgrid system: A hierarchical trust-region multi-agent reinforcement learning approach},
  author={Xu, Xuesong and Xu, Kai and Zeng, Ziyang and Tang, Jiale and He, Yuanxing and Shi, Guangze and Zhang, Tao},
  journal={Applied Energy},
  volume={375},
  pages={123923},
  year={2024},
  publisher={Elsevier}
}

@article{MARL10,
  title={Trust region policy optimisation in multi-agent reinforcement learning},
  author={Kuba, Jakub Grudzien and Chen, Ruiqing and Wen, Muning and Wen, Ying and Sun, Fanglei and Wang, Jun and Yang, Yaodong},
  journal={arXiv preprint arXiv:2109.11251},
  year={2021}
}

@article{MARL12,
  title={Heterogeneous-agent reinforcement learning},
  author={Zhong, Yifan and Kuba, Jakub Grudzien and Feng, Xidong and Hu, Siyi and Ji, Jiaming and Yang, Yaodong},
  journal={Journal of Machine Learning Research},
  volume={25},
  number={32},
  pages={1--67},
  year={2024}
}

\end{document}